\documentclass[aps, pra, twocolumn,show pacs,show keys,superscript address,superscript reference,floatfix]{revtex4-2}
\usepackage[colorlinks=true,citecolor=blue,linkcolor=blue,urlcolor=blue]{hyperref}
\usepackage[sort&compress]{natbib}
\usepackage{graphicx,times}
\usepackage{color}
\usepackage{amssymb}

\usepackage{amsmath}

\usepackage{algpseudocode}
\usepackage{algorithm}
\usepackage{algorithmicx}

\newcommand{\ket}[1]{\vert #1 \rangle}
\newcommand{\bra}[1]{\langle #1 \vert}
\newcommand{\ketbra}[2]{\vert #1 \rangle \langle #2 \vert}

\DeclareMathAlphabet\mathbfcal{OMS}{cmsy}{b}{n}

\begin{document}
\title{Resilient superconducting-element design with genetic algorithms}

\author{F. A. C\'ardenas-L\'opez}
\email{f.cardenas.lopez@fz-juelich.de}
\affiliation{Forschungszentrum J\"ulich GmbH, Peter Gr\"unberg Institute, Quantum Control (PGI-8), 52425 J\"ulich, Germany} 

\author{J. C. Retamal}
\affiliation{Departamento de F\'isica, Universidad de Santiago de Chile (USACH), Avenida V\'ictor Jara 3493, 9170124, Santiago, Chile}
\affiliation{Center for the Development of Nanoscience and Nanotechnology, Estaci\'on Central, 9170124, Santiago, Chile}

\author{Xi Chen}
\affiliation{Instituto de Ciencia de Materiales de Madrid (CSIC),Cantoblanco, E-28049 Madrid, Spain}

\author{G. Romero}
\affiliation{Departamento de F\'isica, Universidad de Santiago de Chile (USACH), Avenida V\'ictor Jara  3493, 9170124, Santiago, Chile}
\affiliation{Center for the Development of Nanoscience and Nanotechnology, Estaci\'on Central, 9170124, Santiago, Chile}

\author{M. Sanz}
\email{mikel.sanz@ehu.es}
\affiliation{Department of Physical Chemistry, University of the Basque Country UPV/EHU, Apartado 644, E-48080 Bilbao, Spain}
\affiliation{EHU Quantum Center, University of the Basque Country UPV/EHU, Bilbao, Spain}
\affiliation{IKERBASQUE, Basque Foundation for Science, Plaza Euskadi, 5, 48009 Bilbao, Spain}
\affiliation{Basque Center for Applied Mathematics (BCAM), Alameda de Mazarredo 14, 48009 Bilbao, Basque Country, Spain}

\begin{abstract}
We present superconducting quantum circuits which exhibit atomic energy spectrum and selection rules as ladder and lambda three-level configurations designed by means of genetic algorithms. These heuristic optimization techniques are employed for adapting the topology and the parameters of a set of electrical circuits to find the suitable architecture matching the required energy levels and relevant transition matrix elements. We analyze the performance of the optimizer on one-dimensional single- and multi-loop circuits to design ladder ($\Xi$) and lambda ($\Lambda$) three-level system with specific transition matrix elements. As expected, attaining both the required energy spectrum and the needed selection rules is challenging for single-loop circuits, but they can be accurately obtained even with just two loops. Additionally, we show that our multi-loop circuits are robust under random fluctuation in their circuital parameters, i.e. under eventual fabrication flaws. Developing an optimization algorithm for automatized circuit quantization opens an avenue to engineering superconducting circuits with specific symmetry to be used as modules within large-scale setups, which may allow us to mitigate the well-known current errors observed in the first generation of quantum processors.
\end{abstract}
\maketitle
\section{Introduction}
Optimization-assisted circuit design is a successful field in modern electronics to engineer resilient circuit topologies and find appropriate circuital parameters matching some given specifications ~\cite{Branin1967,Gupta1981}. For instance, it has been used in microwave engineering to characterize scattering parameters on transmission line resonators~\cite{Bodharamik1971}, representing the ABCD matrix of two-port networks~\cite{Peres2003}, and to design circuital topologies for multi-port networks~\cite{Murray1969}. In circuit design, a commonly used subroutine corresponds to the genetic algorithm (GA) that mimics Darwinian evolution to optimize functions or solve search problems~\cite{Gen.alg.py,IEEE.3.201}. This algorithm has been widely used in modern electronics to fabricate electrical circuits~\cite{GAbook1,GAbook2}, as well as to design protocols for quantum computing and quantum simulation \cite{Phys.Re.Lett.116.230504,Q.4.1,Quan.Sci.tech.4.014007,1812.01032,QML.1.15,Phys.Rev.R.2.033078}. Besides, technological progress in silicon-based lithographic fabrication has allowed building electronic circuits operating at the microwave regime to exhibit quantum behavior when cooled down to millikelvin temperatures. This architecture, termed as superconducting quantum circuits (SC)~\cite{Devoret2004,Devoret2005book,You2005,Clarke2008,Wendin2005,Devoret2013,Kockum2019,Krantz2019,Kjaergaard2020,Martinis2020}, relies on devices made of Josephson junctions for implementing two-level systems, while LC oscillators and transmission lines~\cite{Jour.Appl.Phys.104.113904} serve as the quantized electromagnetic field mode(s). The interplay between these systems spared the field circuit quantum electrodynamics \cite{Phys.Rev.A.69.062320,Nature.431.162,Nature.431.159,Nature.451.664,arXiv.2005.12667,Blais2020} and microwave quantum photonics~\cite{Phot.544,Phys.Rep,Yang2021}. 

Electrical circuits described by a given quantum Hamiltonian~\cite{Nigg2012,Solgun2014,Ulrich2016,1,Parra2019,You2019} exhibit specific spectral and dynamical properties. Engineering the circuit architecture could allow us to modify spectral properties for particular purposes. Several works proposed engineer protocols to design superconducting circuit architectures with a given anharmonicity starting from a predefined circuit architecture~\cite{Yan2020}. On the other hand, some protocols have been used to implement closed-loop optimization to build flux-qubits, outperforming existing proposals and coupling them to create a 4-local coupler~\cite{Menke2020}. Likewise, minimization subroutines have been proposed to create quantum processors based on transmons~\cite{Kyaw2020}. Besides, different circuit analyzers have been presented to study the energy spectrum and circuit properties. Therefore, optimization techniques seem to be appropriate to fabricate and test the next generation of superconducting processors~\cite{Gely2020,Genois2021,Aumann2021}. 

In nature, selection rules due to the dipolar interaction between an atom and a quantized field mode~\cite{Scully} is an essential ingredient in light-matter interaction. These selection rules, fixed by nature, permit us to classify the atomic systems accordingly to the states connected by the interaction, such as $\Xi$ (ladder), $\Lambda$ (lambda), and $V$ (V-type)~\cite{Nature.474.589}. In contrast, circuit QED offers flexibility to engineer selection rules through external signals or choosing appropriate circuit parameters to break the internal symmetry allowing to access energy states differently compared to atomic systems~\cite{PhysRevLett.95.087001,Murali2004,Srinivasan2011,Earnest2018,Vool2018}.

Quantum computation protocols rely on qubits as the basic units. Available quantum hardware consists of weakly-anharmonic systems requiring active control techniques to suppress higher energy levels to maintain the two-level approximation~\cite{PhysRevA.103.032417}. That imposes a tradeoff between anharmonicity and long coherence times. On the other hand, in the last years, there has been an increasing interest in qutrit-based quantum computation. Using qutrits as a basic unit for quantum computation would offer several advantages, such as decreasing the execution time of gated-based algorithm~\cite{PhysRevA.85.062321}, reduced the number of steps needed on a calculation~\cite{Nat.Phys.5.134,Phys.Rev.A.75.022313,Phys.Rev.Lett.94.230502}, avoiding the decoherence effects on the device~\cite{Phys.Rev.A.62.052309,J.Modern Optics.49.2115}, and enhancing the robustness in quantum cryptography protocols~\cite{PhysRevLett.88.127901,PhysRevLett.85.3313,Phys.Rev.Lett.98.060503}. In other contexts, qutrits have been proposed to engineer energy bands for quantum materials~\cite{PhysRevLett.120.130503,PhysRevLett.122.010501}. Exploiting the potential of qutrit-based quantum computation may lead to advantages over the state-of-the-art quantum computation~\cite{Phys.Rev.Lett.105.223601,Phys.Rev.Lett.126.210504,Phys.Rev.X.11.021010,2009.00599}. 

In this article, we propose a bio-inspired design of multi-loop superconducting circuits controlling the energy spectrum and selection rules. Starting from the general Hamiltonian of a closed-loop circuit with four branches, we optimize their positions and circuit parameters to design energy levels satisfying similar spectral and dynamical properties of atomic systems. We illustrate our automatized circuit quantization algorithm to generate resilient three-level systems with ladder $\Xi$ and lambda $\Lambda$ structure. Our results show that for single-loop circuits with two and three branches, there is a \textit{tradeoff} between attaining both the required energy spectrum and the selection rule of the artificial atom. The such tradeoff is not present in circuits built up by a linear array of single-loop circuits in the so-called \textit{multi-loop} circuits, in which we observe that the energy spectrum and the selection rules of the artificial atom do not depend on the control parameter (external magnetic flux). Also, we show that the energy spectrum and the transition matrix elements of the multi-loop circuits are robust under random fluctuations in their circuital parameters. Our findings may pave the way for developing modular quantum systems that exhibit required spectral features that are robust under variations in their parameters employing (bio-inspired) heuristic optimization algorithms.

The manuscript is organized as follows. In Section \ref{II}, we introduce our heuristic optimization algorithm based on the genetic algorithm using individuals single and multi-mode loops containing four links. In section \ref{III}, we describe our automatized circuit quantization algorithm that finds the adequate matrix representation of the circuit Hamiltonian choosing between the charge basis and annihilation and creation bosonic operators, respectively. Section \ref{IV} presents results concerning optimal circuit configurations. In section \ref{V}, we study the dependence of such designs on the external flux. In section \ref{VI}, we analyze the dynamical properties of these optimal configurations. In section \ref{VII}, we consider studying optimal circuit architecture under fluctuations in physical parameters. Finally, we present the conclusions.

\section{Genetic algorithm for adaptive circuit design}\label{II}
\begin{figure}[!b]
\centering
\includegraphics[width=1\linewidth]{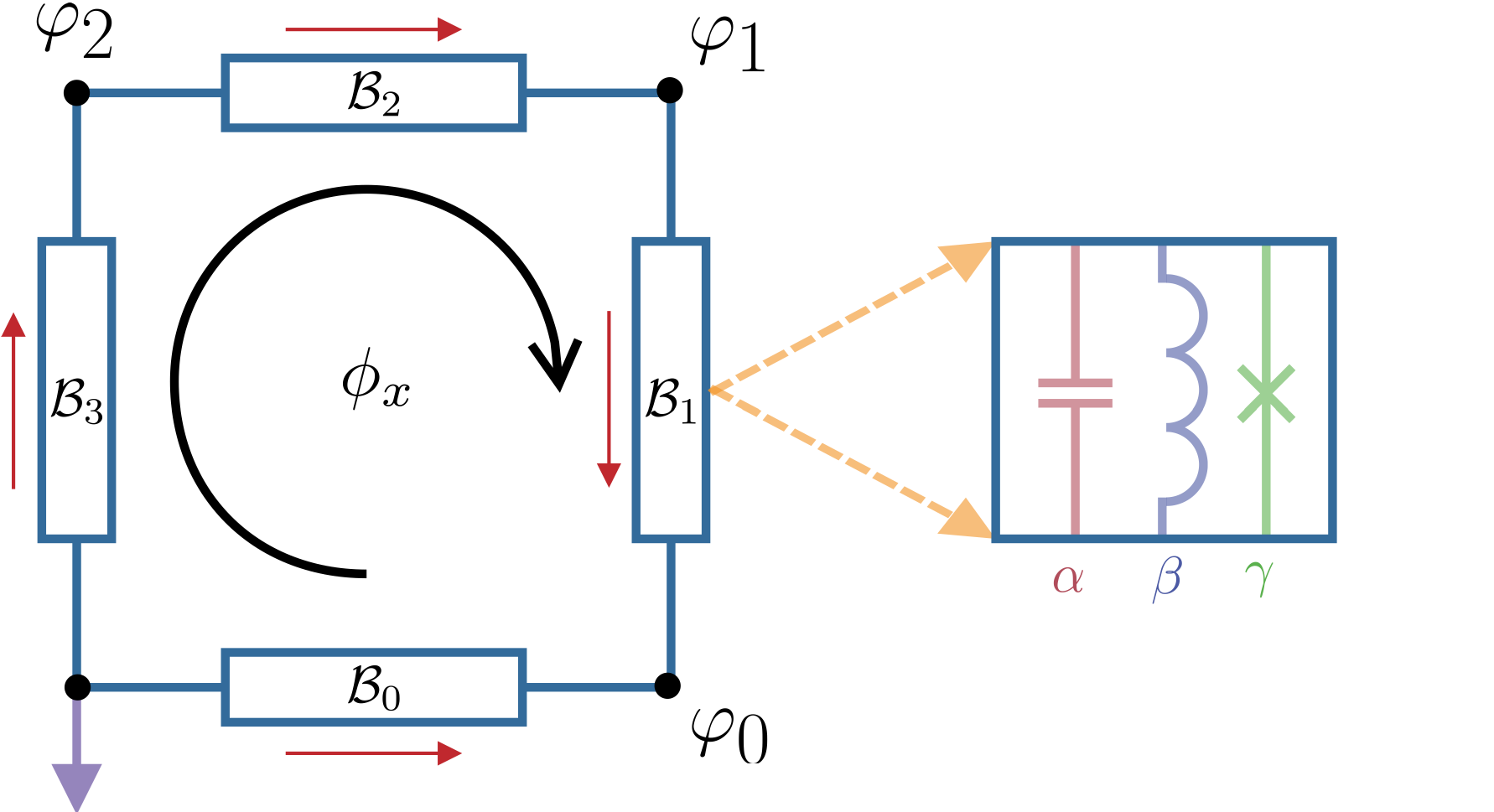}
\caption{(Color online) Schematic illustration of the single-loop circuit formed by four links $\mathcal{B}_{\ell}$ threaded by an external flux $\phi_{x}$. We describe each link as a parallel-connected circuit.}
\label{fig:fig1}
\end{figure}
Genetic algorithms are heuristic optimization algorithms inspired by Darwinian selection. In our problem, the {\it individuals} correspond to circuit configurations consisting of single or multi-loop elements as depicted in Fig.~\ref{fig:fig1}. Different circuit elements lead to distinct individuals. As we aim at engineering architectures satisfying specific energy spectra with given selection rules, these individuals are evaluated and scored depending on how well they craved the required conditions through the {\it cost function}. To find the optimal circuit, a crucial step relies on automatizing the circuit quantization method to compute each individual's spectrum and transition element. In the first part of this section, we focus on developing the tools to perform this automatic diagonalization.\par
Let us consider a single-loop circuit with four links, each containing randomly chosen elements among capacitors, inductors, Josephson junctions, or just their absence. The general configuration of the loop has associated a parallel-connected non-linear circuit to each link $\mathcal{B}_{\ell}$, see Fig.~\ref{fig:fig1}. In this derivation, we have assumed that all the values of the circuit parameters are fixed. The $\ell$th link is characterized by the array $\{\alpha_{\ell},\beta_{\ell},\gamma_{\ell}\}$ that contains all the information about which circuit element corresponds to that link. Thus, the parameters $\{\alpha_{\ell},\beta_{\ell},\gamma_{\ell}\}$ correspond to binary numbers having $\{0,1\}$ as entries, such that only one of them is equal to one and the others are zero. For example, the set $\{1,0,0\}$ refers to the $\ell$th link as a capacitor, whereas the configuration $\{0,0,0\}$ refers to the link absence (wired connection), respectively.

In order to avoid superfluous variables describing the circuit, we represent its Lagrangian in terms of the nodes variables~\cite{1}, which can be conveniently expressed as the following quadratic form  
\begin{eqnarray}
	\label{Lagrangian}\nonumber
	\mathcal{L} &=& \bigg(\frac{1}{2}\bigg)\bigg(\frac{\Phi_{0}}{2\pi}\bigg)^{2}\big[\dot{\vec{\varphi}}^{{\rm{T}}}\hat{C}\dot{\vec{\varphi}} - \vec{\varphi}^{{\rm{T}}}\hat{L}^{-1}\vec{\varphi}\big]\\\nonumber
	&+&\gamma_{0}E_{J_{0}}\cos\big(\varphi_{0}\big)+\gamma_{1}E_{J_{1}}\cos\big(\varphi_{1}-\varphi_{0}\big)\\
	&+&\gamma_{2}E_{J_{2}}\cos\big(\varphi_{2}-\varphi_{1}\big)+\gamma_{3}E_{J_{3}}\cos\big(\varphi_{2}\big).
\end{eqnarray}
Here, $\Phi_{0}=h/(2e)$ is the magnetic quantum flux, with $e$ as the electric charge, moreover, $\vec{\varphi}=\{\varphi_{0},\varphi_{1},\varphi_{2}\}$ corresponds to the phase variable vector describing the $k$th node. Furthermore, $\hat{C}$ and $\hat{L}^{-1}$ are the capacitance and the inverse of the inductance matrix
\begin{eqnarray}
	\label{Eqn6}
	\hat{C} &=& \left(
	\begin{array}{cccc}
		C_{\Sigma_{0}}+C_{\Sigma_{1}} & -C_{\Sigma_{1}} & 0 \\
		-C_{\Sigma_{1}} & C_{\Sigma_{1}}+C_{\Sigma_{2}} & -C_{\Sigma_{2}} \\
		0 & -C_{\Sigma_{2}} & C_{\Sigma_{2}}+C_{\Sigma_{3}}
	\end{array}
	\right),\\
	\label{Eqn7}
	\hat{L}^{-1}&=&\left(
	\begin{array}{cccc}
		\frac{1}{L_{\Sigma_{0}}}+\frac{1}{L_{\Sigma_{1}}} & -\frac{1}{L_{\Sigma_{1}}} & 0  \\
		-\frac{1}{ L_{\Sigma_{1}}} & \frac{1}{L_{\Sigma_{1}}}+\frac{1}{L_{\Sigma_{2}}} & -\frac{1}{ L_{\Sigma_{2}}} \\
		0 & -\frac{1}{ L_{\Sigma_{2}}} &  \frac{1}{L_{\Sigma_{2}}}+\frac{1}{L_{\Sigma_{3}}}
	\end{array}
	\right),
\end{eqnarray}
where, $C_{\Sigma_{\ell}}=(\alpha_{\ell}C_{\ell}+ \gamma_{\ell}C_{J_{\ell}})$ is the equivalent capacitance of the $\ell$th branch. Moreover, $1/L_{\Sigma_{\ell}}=\beta_{\ell}/L_{\ell}$ is the equivalent inductance, where $C_{\ell}$, $L_{\ell}, C_{J_{\ell}}$ and $E_{J_{\ell}}$ represent the capacitance, inductance, Josephson capacitance and Josephson energy of the $\ell$th element.

Notice that it is possible to manipulate the properties of the electrical circuit by threading it with an external magnetic flux $\phi_{x}$. In our Lagrangian, such dependence appears through the fluxoid quantization rule. In our work, we follow the same approach as in Ref.~\cite{You2019}, in which they choose the closure branch where a Josephson junction is placed. In other words, for the array $\vec{\gamma}=\{\gamma_{0},\gamma_{1},\gamma_{2},\gamma_{3}\}$ that contains all the information about the Josephson junction in the circuit, we choose any $\gamma_{\ell}\neq0$ and add $\varphi_x=2\pi\phi_x/\Phi_{0}$ to the corresponding phase difference. In order to automatize this procedure on our algorithm, we select the closure branch as the first one that satisfies $\gamma_{\ell}\neq0$. In our particular case, without loss of generality, we choose the branch ``0'' to apply the fluxoid quantization rule. Thus, the circuit Lagrangian modifies as follows
\begin{eqnarray}
	\label{Lagrangian_closed}\nonumber
	\mathcal{L} &=& \bigg(\frac{1}{2}\bigg)\bigg(\frac{\Phi_{0}}{2\pi}\bigg)^{2}\big[\dot{\vec{\varphi}}^{{\rm{T}}}\hat{C}\dot{\vec{\varphi}} - \vec{\varphi}^{{\rm{T}}}\hat{L}^{-1}\vec{\varphi}\big]\\\nonumber
	&+&\gamma_{0}E_{J_{0}}\cos\big(\varphi_{0}-\varphi_{x}\big)+\gamma_{1}E_{J_{1}}\cos\big(\varphi_{1}-\varphi_{0}\big)\\
	&+&\gamma_{2}E_{J_{2}}\cos\big(\varphi_{2}-\varphi_{1}\big)+\gamma_{3}E_{J_{3}}\cos\big(\varphi_{2}\big).
\end{eqnarray}
Within this approach, it is possible to design several circuit configurations simply by changing the parameters $\vec{\alpha}=\{\alpha_{0},\alpha_{1},\alpha_{2},\alpha_{3}\}$, $\vec{\beta}=\{\beta_{0},\beta_{1},\beta_{2},\beta_{3}\}$, and $\vec{\gamma}=\{\gamma_{0},\gamma_{1},\gamma_{2},\gamma_{3}\}$, respectively. However, we need to discriminate the circuit configurations that may lead to circuits without quantum Hamiltonian as in the case of single-loops containing only capacitors or inductors, respectively. Furthermore, within these configurations, there exist circuits having \textit{passive nodes} where two or mode inductors (capacitors) converge~\cite{1}. In this situation, the circuit will contain more coordinates (velocities) than velocities (coordinates), and these energy terms on the Lagrangian correspond to free particles~\cite{Adrian}. We can eliminate these terms through the Euler-Lagrange equation $\frac{d}{dt}(\partial\mathcal{L}/[\partial[\dot{\varphi}_{\ell}])=\partial\mathcal{L}/\partial[\varphi_{\ell}]$. For the case where a coordinate (velocity) is missing, we obtain that $\partial\mathcal{L}/\partial[\varphi_\ell]=0$ ($\partial\mathcal{L}/\partial[\dot{\varphi}_\ell]=0$). This condition permits us to write the passive node in terms of the active ones. Notice that this procedure is equivalent to calculating the effective inductance (capacitance) on this node. For a detailed explanation of passive node elimination, see Appendix~\ref{A1}.\par
Until here, we obtain a circuit Lagrangian that lacks a superfluous variable since the fluxoid quantization rule and the elimination of the passive nodes. We proceed to calculate the circuit Hamiltonian using the Legendre transformation, obtaining
\begin{eqnarray}
	\label{Hamiltonian_closed}\nonumber
	\mathcal{H} &=& \bigg(\frac{1}{2}\bigg)\big[\vec{q}^{{\rm{T}}}\hat{C}^{-1}\vec{q} + \vec{\varphi}^{{\rm{T}}}\hat{E}_{L}\vec{\varphi}\big]\\\nonumber
	&-&\gamma_{0}E_{J_{0}}\cos\big(\varphi_{0}-\varphi_{x}\big)-\gamma_{1}E_{J_{1}}\cos\big(\varphi_{1}-\varphi_{0}\big)\\
	&-&\gamma_{2}E_{J_{2}}\cos\big(\varphi_{2}-\varphi_{1}\big)-\gamma_{3}E_{J_{3}}\cos\big(\varphi_{2}\big),
\end{eqnarray}
where $\hat{E}_{L}=\Phi_{0}^{2}\hat{L}^{-1}/(2\pi)^{2}$ is the inductive energy matrix. Notice that in superconducting quantum circuits, the electrical charge is proportional to the number of Cooper-pairs on the device. In this way, we can define $\vec{q}=-2e\vec{N}$, where $\vec{N}=\{N_{0},N_{1},N_{2},N_{3}\}$ is the vector describing the number of Cooper-pair associated with each superconducting phase $\vec{\varphi}$. In this representation, we write the circuit Hamiltonian as follows
\begin{eqnarray}
	\label{Hamiltonian_closed_1}\nonumber
	\mathcal{H} &=& 4\vec{N}^{{\rm{T}}}\hat{E}_{C}\vec{N} + \vec{\varphi}^{{\rm{T}}}\frac{\hat{E}_{L}}{2}\vec{\varphi}\\\nonumber
	&-&\gamma_{0}E_{J_{0}}\cos\big(\varphi_{0}-\varphi_{x}\big)-\gamma_{1}E_{J_{1}}\cos\big(\varphi_{1}-\varphi_{0}\big)\\
	&-&\gamma_{2}E_{J_{2}}\cos\big(\varphi_{2}-\varphi_{1}\big)-\gamma_{3}E_{J_{3}}\cos\big(\varphi_{2}\big),
\end{eqnarray}
where $\hat{E}_{C}=e^2\hat{C}^{-1}/2$ is the charge energy matrix.\par
To engineer an electrical circuit that meets desirable specifications, we define a cost function that evaluates all possible candidates to qualify them according to how likely the device fulfills the specifications. This cost function $\mathcal{F}$ takes as input the set $\mathcal{T}=\{\vec{\alpha},\vec{\beta},\vec{\gamma}\}$ that encodes the circuit topology, and the set $\mathcal{P}=\{C_{\ell},L_{\ell}, C_{J_{\ell}},E_{J_{\ell}},C_{c_{\ell}}\}$ that the determines the working regime of the device and provides a real number whose minimum is the craved circuit. In this work, we will first find the optimal circuit topology by fixing the set of circuital parameters (including the external flux). The latter constitutes our minimization unit. Once we find the optimal circuit topology, we search for the optimal set $\mathcal{P}$. The resulting circuit satisfies the desired energy spectrum and transition matrix elements.

\begin{figure}[!t]
	\centering
	\includegraphics[width=1\linewidth]{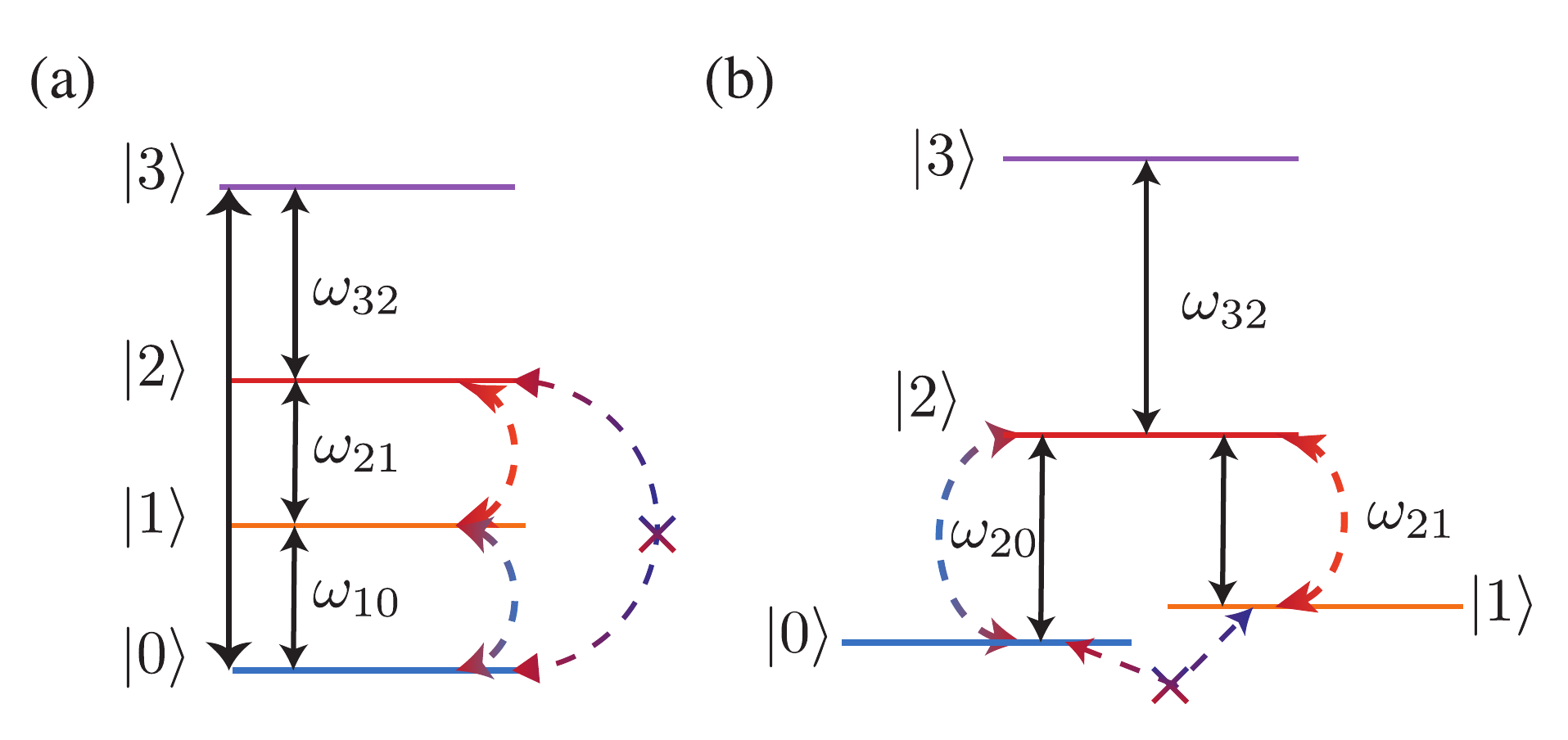}
	\caption{(Color online) Schematic illustration for three-level systems exhibiting (a) a ladder ($\Xi$) configuration and (b) a lambda ($\Lambda$) configuration. Black solid lines represent the transition frequency of the system configuration. Dashed lines correspond to the selection rules observed in the dipolar coupling between natural atoms and a cavity.}
	\label{fig:fig3}
\end{figure}

In our work, we use our automatized circuit quantization method to find the optimal configuration and circuital parameters such that the device works effectively as an atomic system. Here, we will focus on designing a ladder ($\Xi$) and lambda ($\Lambda$) three-level system. For the former, we require that the first two-energy transitions $\omega_{10}=\omega_{1}-\omega_{0}$ and $\omega_{21}=\omega_{2}-\omega_{1}$ be identical, here $\omega_{kj}=\omega_{k}-\omega_{j}$ is difference between the $k$th and the $j$th energy level of the device described by the circuit Hamiltonian. Moreover, to avoid population leakage to higher energy levels, we also demand that the transition $\omega_{32}$ be out of resonance from $\omega_{10}$ and $\omega_{21}$, as depicted in Fig.~\ref{fig:fig3}{(a)}. Finally, to ensure that the energy spectrum is consistent with the experimental implementation of the platform, we bound the low-lying energy spectrum up to a frequency cut-off $\omega_{\rm{M}}=16~[\rm{GHz}]$~\cite{Wendin2005,Rep.Prog.Phys.80.106001,Krantz2019}. We encode these requirements as follows
\begin{eqnarray}\nonumber
	\label{Eqn17}
	d_{1} &=& \left\lvert \omega_{21}-\omega_{10}\right\rvert,d_{2} =\left\lvert \omega_{32}-\omega_{21}\right\rvert-\Gamma,\\
	d_{3} &=& \left\lvert \omega_{32}-\omega_{10}\right\rvert-\Gamma, d_{4} = \left\lvert \omega_{30}-\omega_{\rm{M}}\right\rvert,
\end{eqnarray}
here, $\Gamma$ is a control parameter (with units of frequency) that fixes the detuning between the transitions $\omega_{32}$, $\omega_{10}$ and $\omega_{21}$, respectively. On the other hand, we also require engineering the set of transitions allowed for the device, particularly for the ladder system we demand transitions between adjacent energy levels, such as the states $\ket{0}\leftrightarrow\ket{1}$ and $\ket{1}\leftrightarrow\ket{2}$, and we forbid transitions involving two-photon processes as $\ket{0}\nleftrightarrow\ket{2}$~\cite{Nature.474.589}, where $\ket{k}$ is the $k$th eigenstate of the system. In this direction, we may impose some constraint on the transition matrix elements of the device through the differences  
\begin{eqnarray}\nonumber
	\label{Eqn18}
	d_{5} &=& \left\lvert | \langle0|\hat{\mathcal{O}}|1\rangle|-1 \right\lvert,\quad d_{6}=\left\lvert | \langle1|\hat{\mathcal{O}}|2\rangle|-1 \right\lvert,\\
	d_{7} &=& \left\lvert \langle0|\hat{\mathcal{O}}|2\rangle \right\lvert,
\end{eqnarray}
where $\hat{\mathcal{O}}$ corresponds to a particular operator of the electrical circuit. For instance, we will use the charge operator for our calculations.

Therefore, a good choice for the cost function corresponds to the hypersphere $\mathcal{F}_{\Xi}(\mathcal{T},\mathcal{P})=(1/49)\sum_{i=1}^{7}d_{i}^{2}$.  We apply the same reasoning to define the needed conditions to design a lambda system. Contrary to the ladder, this architecture requires two metastable ground states. Thus, the low-lying energy state of the device should be degenerate, imposing that $\omega_{10}=0$. We also demand that these metastable states have equal energy transitions with the first excited state. In other words, we need that $\omega_{20}=\omega_{21}$, see Fig.~\ref{fig:fig3}{(b)}. These conditions are determined by the differences
\begin{eqnarray}\nonumber
	\label{Eqn19}
	d'_{1} &=& \left\lvert \omega_{10}\right\rvert,d'_{2} =\left\lvert \omega_{20}-\omega_{21}\right\rvert,\\
	d'_{3} &=& \left\lvert \omega_{30}-\omega_{\rm{M}}\right\rvert.
\end{eqnarray}
It is worthwhile noticing that we do not impose the condition $\omega_{32}$ to be out of resonance from $\omega_{20}$ and $\omega_{21}$ to avoid access to higher energy levels. Nonetheless, we will prove later that state $\ket{3}$ does not show up in the system dynamics. Besides, the set of transitions for the lambda system requires no interaction between the two metastable states $\ket{0}\nleftrightarrow\ket{1}$, and only interactions mediated by the first excited state are allowed, i.e., $\ket{0}\leftrightarrow\ket{2}$ and $\ket{1}\leftrightarrow\ket{2}$~\cite{PhysRevLett.95.087001,Nature.474.589}. These conditions are summarized in the following differences
\begin{eqnarray}\nonumber
	\label{Eqn20}
	d'_{4} &=& \left\lvert  \langle0|\hat{\mathcal{O}}|1\rangle\right\lvert,\quad d'_{5}=\left\lvert  \langle0|\hat{\mathcal{O}}|2\rangle|-1 \right\lvert,~~~~~~~~\\	d'_{6} &=& \left\lvert \langle1|\hat{\mathcal{O}}|2\rangle -1\right\lvert.
\end{eqnarray}
Hence, we define the cost function as $\mathcal{F}_{\Lambda}(\mathcal{T},\mathcal{P})=(1/36)\sum_{i=1}^{6}{d'}_{i}^{2}$. Thus, a minimal value of our cost functions will correspond to electrical circuits satisfying the circuit's required spectral and dynamical properties. We carried out the findings of those circuits through two minimization processes. The first one finds the optimal set $\mathcal{T}^{\star}$ that encodes the circuit topology. Here it is convenient to use algorithms that deal with a discrete set since the topology corresponds to arrays of binary numbers. The second minimization relies upon finding the operating regime of the circuit. In other words, we may reach the optimal set $\mathcal{P}^{\star}$ so that the device's spectral and dynamical properties satisfy our requirements and its circuital values be consistent with experimental realization within the quantum platform. This minimization searched the optimal over a continuous set of parameters bounded by the values $C=(0.15,800)~[{\rm{fF}}]$, $L=(238,15000)~[{\rm{pH}}]$, $C_{J}=(3,6000)~[{\rm{fF}}]$, and $E_{J}/h=(0.15,200)~[{\rm{GHz}}]$~\cite{Nat.Phys.16.247,Rep.Prog.Phys.80.106001,arXiv.2005.12667,Pechenezhskiy2020}. 

To obtain the optimal circuit topology ($\mathcal{T}^{\star}$), we need to search over a discrete set (which circuit element to place on a specific branch in the loop) making the Genetic Algorithm (GA)~[6] the most suitable optimization procedure. At this stage of the minimization, we can think of each circuit element as a block and the genetic algorithm creates a set of possible candidates and combines them so that with each iteration, we get closer to a topology that satisfies the requirements imposed by the previously defined cost functions.

For explaining how the GA is implemented, first we should note that the topology of our circuit is completely defined by the matrix $\hat{\mathcal{I}}=(\vec{\alpha}, \vec{\beta}, \vec{\gamma})^{\rm{T}}$, so that the $\ell$th column $\{\alpha_{\ell},\beta_{\ell},\gamma_{\ell}\}$ says which element is on the $\ell$th branch, remember that if all these entries are zero, there is a wiring connection in that branch. In GA language we refer to this matrix as our \textit{individual}. Then, we start by defining our \textit{initial population} consisting in $M$ individuals i.e., $\mathcal{S}=\{\hat{\mathcal{I}}^{(j)},\quad \forall j\in M\}$, and we proceed by calculating the cost function of each of them. In this stage, we select the \textit{parents} that are the $M-p$ configurations with the highest \textit{fitness} or the smallest cost function. Afterwards, we need to generate the \textit{offspring} by mixing them through the \textit{crossover} that is made by randomly choosing two parents $\mathcal{I}^{(j)}$ and $\mathcal{I}^{(k)}$
\begin{eqnarray}
	\begin{pmatrix}\nonumber
		\alpha_{0}^{(j)} & \alpha_{1}^{(j)} & \alpha_{2}^{(j)} & \alpha_{3}^{(j)}\\
		\beta_{0}^{(j)} & \beta_{1}^{(j)} & \beta_{2}^{(j)} & \beta_{3}^{(j)}\\
		\gamma_{0}^{(j)} & \gamma_{1}^{(j)} & \gamma_{2}^{(j)} & \gamma_{3}^{(j)}
	\end{pmatrix},
	\quad \begin{pmatrix}\nonumber
		\alpha_{0}^{(k)} & \alpha_{1}^{(k)} & \alpha_{2}^{(k)} & \alpha_{3}^{(k)}\\
		\beta_{0}^{(k)} & \beta_{1}^{(k)} & \beta_{2}^{(k)} & \beta_{3}^{(k)}\\
		\gamma_{0}^{(k)} & \gamma_{1}^{(k)} & \gamma_{2}^{(k)} & \gamma_{3}^{(k)}
	\end{pmatrix},
\end{eqnarray}
Then, we segregate each parents in subsets of two columns, for example, the individual $j$

\begin{eqnarray}
	\begin{pmatrix}\nonumber
		\alpha_{0}^{(j)} & \alpha_{1}^{(j)} \\
		\beta_{0}^{(j)} & \beta_{1}^{(j)} \\
		\gamma_{0}^{(j)} & \gamma_{1}^{(j)} 
	\end{pmatrix},
	\quad \begin{pmatrix}\nonumber
		\alpha_{2}^{(j)} & \alpha_{3}^{(j)}\\
		\beta_{2}^{(j)} & \beta_{3}^{(j)}\\
		\gamma_{2}^{(j)} & \gamma_{3}^{(j)}
	\end{pmatrix},
\end{eqnarray}
Thus, the offspring are made by combining two columns of the individuals $j$ and $k$, leading to four possible offprints. For example,  a candidate for the offspring will look like
\begin{eqnarray}
	\begin{pmatrix}\nonumber
		\alpha_{0}^{(j)} & \alpha_{1}^{(j)} & \alpha_{0}^{(k)} & \alpha_{1}^{(k)}\\
		\beta_{0}^{(j)} & \beta_{1}^{(j)} & \beta_{0}^{(k)} & \beta_{1}^{(k)}\\
		\gamma_{0}^{(j)} & \gamma_{1}^{(j)} & \gamma_{0}^{(k)} & \gamma_{1}^{(k)}
	\end{pmatrix}
\end{eqnarray}
The next step is to check if the generated offspring corresponds to a valid topology. We discard the offspring if it only contains capacitors, inductors, or oscillator-like configurations. In the first two cases, it is impossible to define a quantum Hamiltonian due to the absence of one of the canonical conjugate variables (phase or charge). In the latter case, the system will have harmonic energy levels, which are not those required by the cost function. Furthermore, if the offspring passes this criterion, we need to regularize the circuit, i.e., find and eliminate passive nodes following the procedure described in Appendix A1. We repeat this process until we obtain $p$ offspring, as we require that the population size remains constant during the optimization process.

To avoid getting stuck in local minima, we introduce \textit{mutations}, which in our context involves randomly selecting one individual from the new population (parents and offspring) and changing one of its columns. For example, in the case of the previous newborn candidate, we may apply a mutation to its third column.
\begin{eqnarray}
	\begin{pmatrix}\nonumber
		\alpha_{0}^{(j)} & \alpha_{1}^{(j)} & 0 & \alpha_{1}^{(k)}\\
		\beta_{0}^{(j)} & \beta_{1}^{(j)} & 0 & \beta_{1}^{(k)}\\
		\gamma_{0}^{(j)} & \gamma_{1}^{(j)} & 0 & \gamma_{1}^{(k)}
	\end{pmatrix}
\end{eqnarray}
This mutation is equivalent to eliminating the circuit element in the third branch and adding a direct wiring connection. The allowed types of mutations include replacing one circuit element with another, or removing an element entirely. It is important to note that mutations at this stage can have a profound impact on the circuit exploration process. For instance, replacing a capacitor with either an inductor or a Josephson junction may result in an architecture with entirely different properties. In this context, mutations are responsible for escaping local minima. At the end of this stage, the newly generated population becomes the initial one for the next iteration and the process is repeated for a fixed number of epochs. The optimal individual $\mathcal{I}^{(\rm{opt})}$ is then selected as the one with the lowest cost function value in the set $\mathcal{S}$, as illustrated in the pseudocode in Algorithm~\ref{A1}. The use of minimization subroutines rather than brute force search relies on the size of the space parameter. The circuit topology space $\mathcal{T}$ for a single loop contains $4^{4}=256$ possible configurations; however, several of them lead to circuits with no Hamiltonian. For instance, electrical circuits with only capacitors or inductors in all links lack a Hamiltonian. We eliminate these configurations from the set, obtaining a parameter space for the single-loop circuit to be $\rm{NC}=4^{4}-\sum_{i=1}^{4}\mathcal{C}_{i}^{4}=225$, where $\mathcal{C}_{i}^{n}$ is the combinatorial symbol. For the one-dimensional array having $N$ sites, as the depicted in Fig.~3, the number of combination scales as $\rm{NC}_{\rm{array}}={\rm{NC}}^{N}$, respectively.\par
\begin{algorithm}[H]
	\caption{Adaptive circuit design algorithm}
	\label{GA}
	\begin{algorithmic}[1]
		\State Minimization($\mathcal{T}$,$\mathcal{P}$)
		\Comment{Fix circuit parameters and change the topology}
		\State Define the initial population of $M$ multi-loops
		\For{$n$ in epoch}
		\State Calculate the fitness ($\mathcal{F}_{\Xi}$,$\mathcal{F}_{\Lambda}$) of each individual using Algorithm~\ref{CQ}
		\State Choose the $M-p$ individuals with the highest fitness
		\State \textbf{Crossover:}
		\For {$i$ in $M-p$}
		\State Divide the circuit using its active nodes $\varphi$
		\State Merge two of them randomly
		\If {the merged circuit is a valid configuration}
		\State Save it as an offspring
		\Else
		\State Repeat
		\EndIf
		\EndFor
		\State \textbf{Mutation:}
		\If {random number $\leq$ 0.01}
		\State Change a circuit element of the individual
		\EndIf
		\State Update the initial population with the offsprings
		\EndFor
		\State \textbf{Optimal circuit with fixed parameters $\mathcal{P}$}
		\Comment{Fix the topology and change the circuit parameters}
		\State Define the circuit parameters for the $M$ multi-loops
		\For{$n$ in epoch}
		\State Calculate the fitness ($\mathcal{F}_{\Xi}$,$\mathcal{F}_{\Lambda}$) of each individual using Algorithm~\ref{CQ}
		\State Choose the $M-p$ individuals with the highest fitness
		\State \textbf{Crossover:}
		\For {$i$ in $M-p$}
		\State Divide the array containing the circuit parameters
		\State Concatenate two of them randomly
		\State Save the result in the offsprings array
		\EndFor
		\State \textbf{Mutation:}
		\If {random number $\leq$ 0.01}
		\State Change the value of a random circuit parameter
		\EndIf
		\State Update the initial population with the offsprings
		\EndFor
		\State \textbf{Optimal circuit}
	\end{algorithmic}
\end{algorithm}

After finding the circuit topology, we need to search its optimal parameters $\mathcal{P}^{\star}$. Notice that for finding the optimal circuit parameters we have more freedom in choosing different minimization method because of the continuous nature of the parameter space. We have selected GA to be consistent with the method employed in the previous step but also because GA algorithm is less likely to get stuck in local minima in comparison with gradient-based and sampling optimization subroutines (see Fig.~\ref{fig:figA4} for a comparison with different approaches). In this part of the optimization, the individual will be a linear array containing all the circuit parameters which are concatenated by branches, explicitly, the individual $\mathcal{I}=\{\mathcal{P}_{0}, \mathcal{P}_{1}, \mathcal{P}_{2}, \mathcal{P}_{3}\}$, where $\mathcal{P}=\{C_{\ell},L_{\ell}, C_{J_{\ell}},E_{J_{\ell}}\}$ are the circuital parameters of each branch. For not biasing the optimization procedure, we always initialize the circuit parameters in one of the borders of the landscape i.e, we select the lower/upper bound of the circuit parameters. The crossover in that case works similar to the discrete optimization; we divide the individual in two parts and create the offspring mixing them e.g., $\mathcal{I}^{(\rm{new})}=\{\mathcal{P}_{2}^{(j)}, \mathcal{P}_{3}^{(j)}, \mathcal{P}_{2}^{(k)}, \mathcal{P}_{3}^{(k)}\}$. We introduce mutation on the optimization by randomly selecting one branch $\mathcal{P}_{\ell}^{(k)}$, let us assume that we have a capacitor $C_{\ell}$ and  we change its value by $C_{\ell}\rightarrow C_{\ell}+(-1)^{r}\mathcal{R}(C_{\rm{min}}, C_{\rm{max}})$, where $r$ is a random binary parameter so that we increase or decrease the current value of the circuit element, and $\mathcal{R}(C_{\rm{min}}, C_{\rm{max}})$ is a random number between the range of the exploration. This optimization is computationally less demanding than the topology ones since the circuital parameter space for a single-loop circuit has dimension eight at most, corresponding to a device with four Josephson junctions. Other single-loop configurations will have fewer parameters. Thus, for a one-dimensional array having $N$ sites, the number of parameters scales as $9N-1$. In 1 we have written a \textit{pseudo-code} of our algorithm.

In our simulations, we selected 16 random circuits as the initial population for the topology optimization. On the other hand, for the multi-loop circuit, we have chosen 16 random arrays of different combinations of the optimal circuit arrangement obtained during the single-loop optimization. For both minimization, we have used a mutation probability of $0.01\%$.
\section{Automatized circuit quantization}\label{III}
Thus far, we have described how to engineer the topology and choose the adequate circuital parameters using a bio-inspired algorithm such that the resulting electrical circuit satisfies the same energy spectrum and selection rules as atomic three-level systems. The genetic algorithm proposed here requires the Hamiltonian matrix representation to compute the previously defined cost function. Thus, choosing the correct quantization basis is mandatory so that the electrical circuits encompass all the underlying physics. The starting point of our automatized circuit quantization method is the classical Hamiltonian given in Eq.~(\ref{Hamiltonian_closed_1})
\begin{eqnarray}\nonumber
	\mathcal{H} &=& 4\vec{N}^{{\rm{T}}}\hat{E}_{C}\vec{N} + \vec{\varphi}^{{\rm{T}}}\frac{\hat{E}_{L}}{2}\vec{\varphi}\\\nonumber
	&-&\gamma_{0}E_{J_{0}}\cos\big(\varphi_{0}-\varphi_{x}\big)-\gamma_{1}E_{J_{1}}\cos\big(\varphi_{1}-\varphi_{0}\big)\\\nonumber
	&-&\gamma_{2}E_{J_{2}}\cos\big(\varphi_{2}-\varphi_{1}\big)-\gamma_{3}E_{J_{3}}\cos\big(\varphi_{2}\big).
\end{eqnarray}
We obtain the quantum Hamiltonian by promoting the number of Cooper-pair and phase variables as quantum operators satisfying canonical commutation relation. In our work, the circuit topology determines the quantization basis for each node operator; for devices only having capacitors and Josephson junctions, we call them Cooper-pair-box-like configurations. In this case, $\hat{N}=\sum N\ketbra{N}{N}$ is a diagonal operator, and the cosine term represents tunneling of Cooper pairs, and they are expressed in terms of $\exp(i\hat{\varphi})\equiv \sum\ket{N+1}\bra{N}$  which satisfy the canonical commutation relations in the form $[\hat{N},\exp(\pm i\hat{\varphi})]=\exp(\pm i\hat{\varphi})$. On the other hand, it is also possible that our circuit provides configurations only having capacitors and inductors or capacitors, inductors, and Josephson junctions as the case of the Fluxonium circuit~\cite{,Nguyen2019}. In these cases, the quantization basis corresponds to the Fock basis where we define annihilation and creation bosonic operators $a$ and $a^{\dag}$ that satisfy the commutation relation $[a,a^{\dag}]=1$.
\begin{figure}[!t]
	\centering
	\includegraphics[width=1\linewidth]{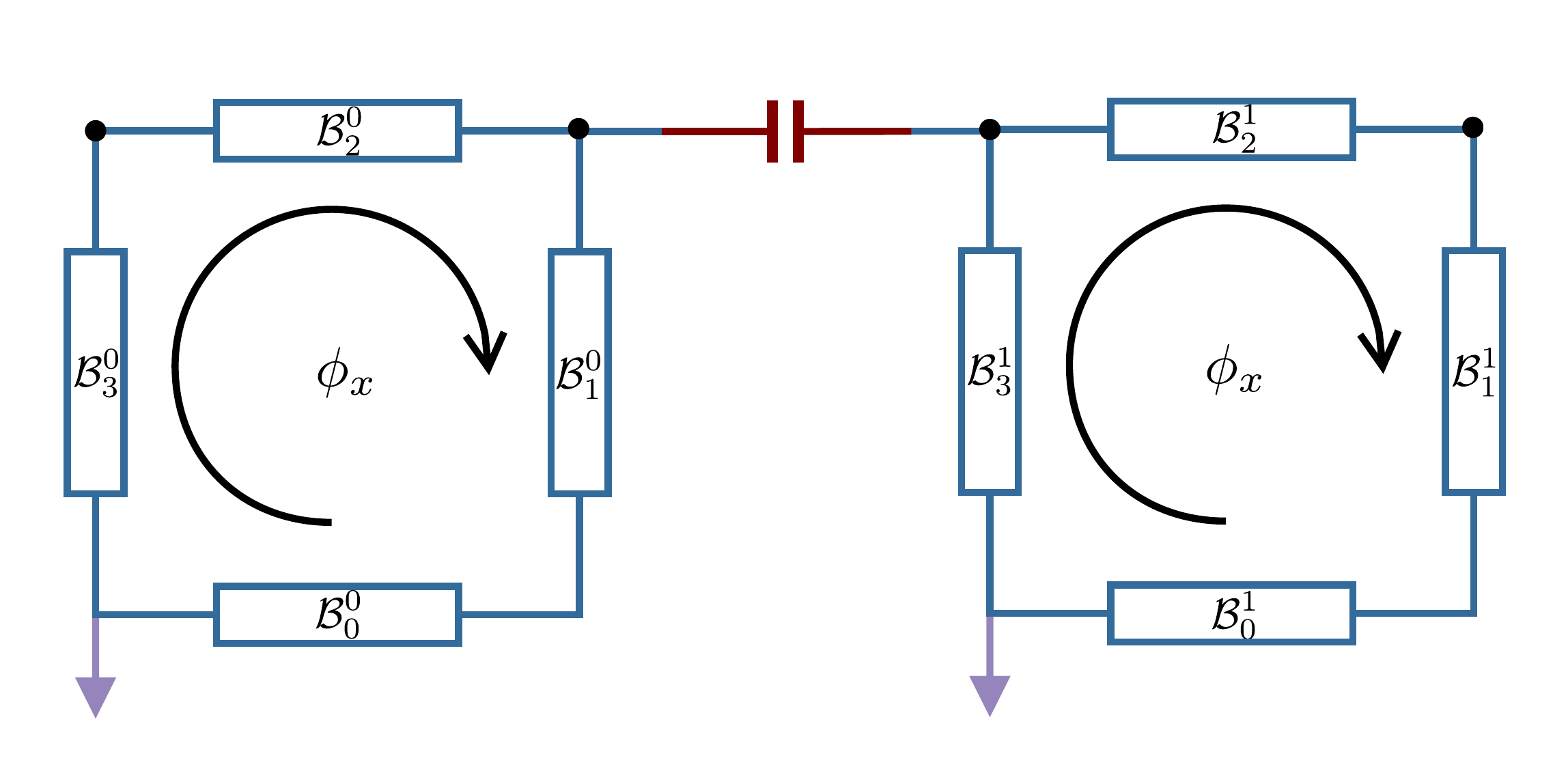}
	\caption{(Color online) Schematic illustration of a multi-loop device in a one-dimensional array consisting of two single-loop circuits with four links threaded by an external magnetic flux. The single loops couple to capacitors through each subsystem's left and right nodes, respectively.}
	\label{fig:fig2}
\end{figure}
To automatize the selection of the quantization basis for each node, we have to look more in depth at the potential term of the circuit Hamiltonian in Eq.~\ref{Hamiltonian_closed_1}. For this analysis, we expand the potential up to the fourth order in the phases $\vec{\varphi}$ ($\varphi_{x}=0$)
\begin{eqnarray}
	\label{U_closed}\nonumber
	\mathcal{U}^{(4)}(\vec{\varphi}) &=& \vec{\varphi}^{{\rm{T}}}\bigg(\frac{\hat{E}_{L}}{2}\bigg)\vec{\varphi}\\\nonumber
	&-&(\gamma_{0}E_{J_{0}}+\gamma_{1}E_{J_{1}}+\gamma_{2}E_{J_{2}}+\gamma_{3}E_{J_{3}})\\\nonumber
	&+&\frac{\gamma_{0}E_{J_{0}}}{2}\varphi_{0}^2+\frac{\gamma_{1}E_{J_{1}}}{2}(\varphi_{0}-\varphi_{1})^2\\\nonumber
	&+&\frac{\gamma_{2}E_{J_{2}}}{2}(\varphi_{1}-\varphi_{2})^2+\frac{\gamma_{3}E_{J_{3}}}{2}\varphi_{2}^2\\\nonumber
	&-&\frac{\gamma_{0}E_{J_{0}}}{24}\varphi_{0}^4-\frac{\gamma_{1}E_{J_{1}}}{24}(\varphi_{0}-\varphi_{1})^4\\
	&-&\frac{\gamma_{2}E_{J_{2}}}{24}(\varphi_{1}-\varphi_{2})^4-\frac{\gamma_{3}E_{J_{3}}}{24}\varphi_{2}^4.
\end{eqnarray}
advert that the expansion coefficients are functions of the Josephson energies. Afterward, we calculate the second and fourth order derivative of the expanded potential $\mathcal{U}^{(4)}(\vec{\varphi})$ with respect to $\varphi_{\ell}$, as an example, we will consider the phase $\varphi_{0}$
\begin{eqnarray}
	\label{derivative_U_2}\nonumber
	\frac{\partial^{2}\mathcal{U}^{(4)}(\vec{\varphi})}{\partial \varphi_{0}^2}&=&[\hat{E}_{L}]_{0,0}+\gamma_{0}E_{J_{0}}-\gamma_{1}E_{J_{1}}+\gamma_{1}E_{J_{1}}\varphi_{0}\varphi_{1}\\\nonumber
	&-&\frac{(\gamma_{0}E_{J_{0}}+\gamma_{1}E_{J_{1}})}{2}\varphi_{0}^2+\gamma_{1}E_{J_{1}}\varphi_{0}\varphi_{1}\\\nonumber
	&-&\frac{\gamma_{1}E_{J_{1}}}{2}\varphi_{1}^2+\frac{\gamma_{1}E_{J_{1}}}{4}\varphi_{0}^2\varphi_{1}^2-\frac{\gamma_{1}E_{J_{1}}}{6}\varphi_{0}\varphi_{1}^3\\
	&+&\frac{\gamma_{1}E_{J_{1}}}{24}\varphi_{1}^4-\frac{\gamma_{1}E_{J_{1}}}{48}\varphi_{0}^2\varphi_{1}^4,\\\nonumber
	\label{derivative_U_4}
	\frac{\partial^{4}\mathcal{U}^{(4)}(\vec{\varphi})}{\partial \varphi_{0}^4}&=&-\gamma_{0}E_{J_{0}}-\gamma_{1}E_{J_{1}}+\frac{\gamma_{1}E_{J_{1}}}{2}\varphi_{1}^{2}-\frac{\gamma_{1}E_{J_{1}}}{24}\varphi_{1}^{4}.\\
\end{eqnarray}
where $[\hat{E}_{L}]_{0,0}$ is the matrix element of the inductance energy matrix. From the equations Eq.~(\ref{derivative_U_2}) and Eq.~(\ref{derivative_U_4}) it is possible to define the criteria to choose the charge or the oscillator basis quantization. If $\partial^{4}\mathcal{U}^{(4)}(\vec{\varphi})/\partial \varphi_{0}^4$ vanishes, but $\partial^{2}\mathcal{U}^{(4)}(\vec{\varphi})/\partial \varphi_{0}^2$ does not it means that there is no cosine potential associated with node variable $\varphi_{0}$. Consequently, we quantize the phase and its corresponding charge using the annihilation and creation operator of the quantum harmonic oscillator. On the contrary, we may think that if $\partial^{4}\mathcal{U}^{(4)}(\vec{\varphi})/\partial \varphi_{0}^4$ and $\partial^{2}\mathcal{U}^{(4)}(\vec{\varphi})/\partial \varphi_{0}^2$ are not zero, the node variable $\varphi_{0}$ could be quantized in terms of the charge basis. However, this is not in general true. For instance, a fluxonium-like configuration consisting of capacitors, inductance, and Josephson junctions fulfills the above-mentioned conditions. However, this architecture is usually quantized in terms of annihilation and creation bosonic operators. To circumvent this problem, we add the constraint $\vec{\beta}[\ell]=0$, meaning that the $\ell$th branches have not associated inductance and consequently. If a given node meets these three conditions, we quantize this degree of freedom using the charge basis. This procedure allows us to automatize the quantization basis of each node using physical arguments regarding in the presence/absence of Josephson junction and inductors on the electrical circuits. We repeat this procedure for all the active nodes on the circuit.\par

For a circuit only containing capacitors and inductors, our automatized method for choosing the quantization basis leads to the following quantum Hamiltonian
\begin{eqnarray}
	\label{H_osc}
	\mathcal{H}=4\vec{N}^{~T}\hat{E}_{C}\vec{N} + \frac{1}{2}\vec{\varphi}^{~T}\hat{E}_{L}\vec{\varphi},
\end{eqnarray}
this Hamiltonian describes a set of coupled harmonic oscillators that can be quantized using the operators $\hat{N}_{\ell}$ and $\hat{\varphi}_{\ell}$ defined in terms of annihilation and creation bosonic oscillator as
\begin{eqnarray}
	\label{Eqn10}\nonumber
	\hat{N_{\ell}} = \sqrt[4]{\frac{[\hat{E}_{L}]_{\ell,\ell}}{32[\hat{E}_{C}]_{\ell,\ell}}}i(a^{\dag}_{\ell}-a_{\ell}),\quad \hat{\varphi_{\ell} }=\sqrt[4]{\frac{2[\hat{E}_{C}]_{\ell,\ell}}{[\hat{E}_{L}]_{\ell,\ell}}}(a^{\dag}_{\ell}+a_{\ell}).
\end{eqnarray}
Here, $[\hat{E}_{C}]_{\ell,\ell}$, and $[\hat{E}_{L}]_{\ell,\ell}$ are the diagonal elements of the charge and inductive matrix, respectively. These operators satisfy canonical commutation relations $[\hat{\varphi}_{\ell},\hat{n}_{\ell'}]=i\hbar\delta_{\ell,\ell'}$ that leads to the Hamiltonian of single-loop circuit ($\hbar=1$)
\begin{eqnarray}
	\label{Eqn11}\nonumber
	\mathcal{H}&=&\sum_{\ell=0}^{2}\omega_{\ell}a_{\ell}^{\dag}a_{\ell} +  \sum_{\ell\neq\ell'}\mathcal{C}_{\ell,\ell'}(a^{\dag}_{\ell}-a_{\ell})(a^{\dag}_{\ell'}-a_{\ell'})\\\nonumber
	&+&\sum_{\ell\neq\ell'}\mathcal{G}_{\ell,\ell'}(a^{\dag}_{\ell}+a_{\ell})(a^{\dag}_{\ell'}+a_{\ell'}).
\end{eqnarray}
where $\omega_{\ell}$, $\mathcal{C}_{\ell,\ell'}$ and $\mathcal{G}_{\ell,\ell'}$ correspond to the oscillator frequency, capacitive and inductive coupling strength given by
\begin{subequations} 
\begin{eqnarray}
	\label{Eqn12a}
	\omega_{\ell}&=&\sqrt{8[\hat{E}_{C}]_{\ell,\ell}[\hat{E}_{J}]_{\ell,\ell}},\\
	\label{Eqn12b}
	\mathcal{C}_{\ell,\ell'}&=&-\frac{[\hat{E}_{C}]_{\ell,\ell'}}{\sqrt{[\hat{E}_{C}]_{\ell,\ell}[\hat{E}_{C}]_{\ell',\ell'}}}\sqrt{\omega_{\ell}\omega_{\ell'}},\\
	\label{Eqn12c}
	\mathcal{G}_{\ell,\ell'}&=&\frac{[\hat{E}_{L}]_{\ell,\ell'}}{\sqrt{[\hat{E}_{L}]_{\ell,\ell}[\hat{E}_{L}]_{\ell',\ell'}}}\sqrt{\omega_{\ell}\omega_{\ell'}}.
\end{eqnarray}
\end{subequations}
On the other hand, for a circuit with two nodes only containing capacitors and Josephson junctions, the automatized method gives us the following circuit Hamiltonian
\begin{eqnarray}
	\label{H_jun}\nonumber
	\mathcal{H}&=&4\vec{N}^{~T}\hat{E}_{C}\vec{N}-\gamma_{0}E_{J_{0}}\cos\big(\hat{\varphi}_{0}-\varphi_{x}\big) \\
	&-&\gamma_{1}E_{J_{1}}\cos\big(\hat{\varphi}_{1}-\hat{\varphi}_{0}\big).
\end{eqnarray}
In the charge basis, the $\hat{N}$ and cosine operators are expressed in the following form
\begin{eqnarray}
	\hat{N}_{\ell}=\sum_{N_{\ell}}^{\infty} N_{\ell}\ketbra{N_{\ell}}{N_{\ell}},~ \exp(i\hat{\varphi})= \sum_{N_{\ell}}^{\infty}\ket{N_{\ell}+1}\bra{N_{\ell}}.
\end{eqnarray}
These operators satisfies the canonical commutation relation $[\hat{N}_{\ell},\exp(\pm i\hat{\varphi}_{\ell'})]=\delta_{\ell,\ell'}\exp(\pm i\hat{\varphi}_{\ell})$. The quantum circuit Hamiltonian reads
\begin{eqnarray}
\label{H_jun_2}
\mathcal{H}&=&4[\hat{E}_{C}]_{0,0}\hat{N}_{0}^{2} + 4[\hat{E}_{C}]_{1,1}\hat{N}_{1}^{2} + 8[\hat{E}_{C}]_{0,1}\hat{N}_{0}\hat{N}_{1}\\\nonumber
&-&\frac{\gamma_{0}E_{J_{0}}}{2}\big[e^{i\varphi_{x}}\ket{N_{0}+1}\bra{N_{0}}+e^{-i\varphi_{x}}\ket{N_{0}}\bra{N_{0}+1}\big]\\\nonumber
&-&\frac{\gamma_{1}E_{J_{1}}}{2}\bigg[\ket{N_{0}+1,N_{1}}\bra{N_{0},N_{1}+1}+{{\rm{H.c}}}\bigg].
\end{eqnarray}
Therefore, these two examples demonstrate the relevance of circuit topology in the representation of the circuit operator, leading to different energy spectra and selection rules. In this context, since the relevant condition in our cost functions relies on having an anharmonic energy spectrum, we can pre-process the possible topologies by discarding all circuit topologies that contain only capacitors and inductors, as their energy spectra are harmonic.  We extend the automatized quantization procedure to one-dimensional array of $N$ single-loop capacitive connected to each other  [see Fig. \ref{fig:fig2}] described through the Hamiltonian
\begin{eqnarray}
\label{Eqn16}
\mathcal{H}_{\mathcal{A}} &=& \sum_{k=0}^{N}\mathcal{H}^{k}+ \sum_{k=0}^{N-1}4E_{C_{c}}\hat{N}_{k}\hat{N}_{k+1},
\end{eqnarray} 
\begin{algorithm}[H]
\caption{Circuit quantization subroutine}
\label{CQ}
\begin{algorithmic}[1]
\State For the defined topology $\mathcal{T}$ of $M$ multi-loop circuits
\For{$j$ in $M$}
	\State calculate the circuit Lagrangian $\mathcal{L}$
	\State \textbf{delete the passive nodes}
\For{$k$ in number of variables $\varphi_k$}
\State compute $\frac{\partial \mathcal{L}}{\partial[\dot{\varphi}_{k}]}=0$ and $\frac{\partial \mathcal{L}}{\partial[\varphi_{k}]}=0$
	\State \emph{solve for the coordinate/velocity missing}
	\State replace $\mathcal{L} \gets \{\varphi_k,\dot{\varphi}_k\}$
\EndFor 
\State $\textit{Hamiltonian $\mathcal{H}=\sum_{k}q_k\varphi_k-\mathcal{L}$}$
\State \textbf{Automatized circuit quantization}
\For{$\ell$ in number of variables $\varphi_\ell$}
\State compute $\frac{\partial^{2}\mathcal{U}^{(4)}(\vec{\varphi})}{\partial \varphi_{\ell}^2}$, $\frac{\partial^{4}\mathcal{U}^{(4)}(\vec{\varphi})}{\partial \varphi_{\ell}^4}$ and $\vec{\beta}[\ell]$
\If{$\frac{\partial^{4}\mathcal{U}^{(4)}(\vec{\varphi})}{\partial \varphi_{\ell}^4}=0$}
\State $q_{\ell}\propto i(a_{\ell}^\dag - a_{\ell})$ and $\varphi_{\ell}\propto (a_{\ell}^\dag + a_{\ell})$
\ElsIf{$\frac{\partial^{4}\mathcal{U}^{(4)}(\vec{\varphi})}{\partial \varphi_{\ell}^4}!=0$ \textbf{and} $\vec{\beta}[\ell]!=0$}
\State $q_{\ell}\propto i(a_{\ell}^\dag - a_{\ell})$ and $\varphi_{\ell}\propto (a_{\ell}^\dag + a_{\ell})$
\ElsIf{$\frac{\partial^{4}\mathcal{U}^{(4)}(\vec{\varphi})}{\partial \varphi_{\ell}^4}!=0$ \textbf{and} $\vec{\beta}[\ell]=0$}
\State $q_{\ell}=-2e\sum\ketbra{N}{N}$
\State $\exp(i\varphi_\ell)=\sum\ketbra{N+1}{N}$
\EndIf
\EndFor 
\State \textit{diagonalize the Hamiltonian}
\State \textit{compute the matrix elements}
\EndFor
\end{algorithmic}
\end{algorithm}
where $\mathcal{H}^{k}$ is the single-loop Hamiltonian in Eq. (\ref{Hamiltonian_closed}) for each loop. Furthermore, each site is coupled with a capacitor with strength $E_{C_{c}}=e^2/C_{c}$, where $C_{c}$ is the coupling capacitance. Moreover, $\hat{N}_{k}$ and $\hat{N}_{k+1}$ corresponds to the charge operators of the $k$th and $(k+1)$th single-loop boxes.\par

Exact diagonalization of Hamiltonian~(\ref{Eqn16}) is computationally challenging even for small one-dimensional arrays. This difficulty arises from the Hilbert space dimension, which increases with the number of nodes $\{\hat{\varphi}, \hat{N}\}$ and loops (e.g., the circuit in Fig.~\ref{fig:fig1}). Specifically, a system of $N$ loops each with $M$ modes of size $d$ has a Hilbert space dimension of $\dim(\mathcal{H})=d^{NM}$. To find a single computational unit satisfying the cost function and exhibiting robustness against circuit parameter fluctuations, we focus on a pair of single-loop circuits as shown in Fig.~\ref{fig:fig3}. Note that this scaling differs from simulating a full quantum processor where the number of units must be considered. For a QPU with $D$ units, the total Hilbert space dimension scales as $\dim(\mathcal{H}_{\rm{QUP}})=d^{DNM}$. Alternatives to reduce the computational cost of eigendecomposition include hierarchical diagonalization~\cite{scqubits}, where the Hamiltonian is decomposed into weakly coupled sub-blocks that are diagonalized independently saving memory. Subsequently, interactions are expressed in this new basis and the full Hamiltonian is diagonalized in that representation. Another potential approach involves mapping complex architectures to linear arrays and use tensor network methods to get the system properties. However, as before, weak interactions are necessary for the Density Matrix Renormalization Group (DMRG) condition to hold. Algorithm~\ref{CQ} presents the automated circuit quantization subroutine.
 
\section{Optimal circuit configuration}\label{IV}
We analyze the resulting circuits obtained by minimizing the cost function provided in the previous section. Here, we will focus on the ladder ($\Xi$) and lambda ($\Lambda$) configuration containing two and three links and then extend to an architecture consisting of two coupled single-loop circuits as depicted in Fig.~\ref{fig:fig2}. The circuit topology, circuital parameters, and the energy transition of each optimal circuit for the ladder and lambda three-level system are summarized in tables~\ref{Table1} and \ref{Table2}, respectively. Whereas we provide a graphical illustration of them in Fig.~\ref{fig:figA4}.\par
\begin{table}[t]
\centering
\scalebox{0.9}{
\begin{tabular}{|c|l|l|l|l|l|l|l|}
\hline
\multicolumn{1}{|l|}{\begin{tabular}[c]{@{}l@{}}$(\mathcal{T}^{\star},\mathcal{P}^{\star})$\end{tabular}} & $\alpha$ & $\beta$ & $\gamma$ & $C~\rm{[pF]}$ & $L~\rm{[nH]}$ & $C_{J}~\rm{[pF]}$ & $E_{J}/\hbar~\rm{[GHz]}$ \\ \hline
\multicolumn{8}{|c|}{\textbf{I}}                                             \\ \hline
0 & 0 & 0 & 1 & 0             & 0            & 0.00192            & 1.000              \\ \hline
1 & 0 & 0 & 0 & 0             & 0            & 0	        & 0              \\ \hline
2 & 0 & 0 & 0 & 0 	       		&0             &             &  \\ \hline
3 & 0 & 0 & 1 & 0             & 0            & 0.0542            & 9.127              \\ \hline
\multicolumn{8}{|c|}{\textbf{II}}                                           \\ \hline
0 & 0 & 0 & 1 & 0  & 0            & 1.50            & 88.2              \\ \hline
1 & 0 & 0 & 1 & 0             & 0 & 3.403           & 100              \\ \hline
2 & 0 & 0 & 0 & 0             & 0            & 0 & 0  \\ \hline
3 & 1 & 0 & 0 & 0.00015             & 0            & 0 & 0  \\ \hline
\multicolumn{8}{|c|}{\textbf{III}}                                             \\ \hline
\multicolumn{8}{|c|}{\textbf{Box 1}}                                             \\ \hline
0 & 1 & 0 & 0 & 0.101             & 0            & 0            &0              \\ \hline
1 & 0 & 0 & 1 & 0             & 0            & 1.59& 1              \\ \hline
2 & 0 & 1 & 0 & 0 &18.25       & 0            &0   \\ \hline
3 & 0 & 1 & 0 & 0             & 150            & 0          &             \\ \hline
\multicolumn{8}{|c|}{\textbf{Box 2}}                                            \\ \hline
0 & 0 & 0 & 0 & 0             & 0            & 0  & 0 \\ \hline
1 & 0 & 0 & 0 & 0             & 0 & 0            & 0              \\ \hline
2 & 0 & 0 & 1 & 0             & 0& 0.52            & 59.45              \\ \hline
3 & 0 & 0 & 0 & 0             & 0            & 0            & 0              \\ \hline
\multicolumn{8}{|c|}{\textbf{Coupling}}                                          \\ \hline
0 & 1 & 0 & 0 & 0.803  & 0            & 0           & 0             \\ \hline
\multicolumn{8}{|c|}{\textbf{System parameters }}                                           \\ \hline
 &  &  &  & $\omega_{10}$  & $\omega_{21}$           & $\omega_{32}$           & $\omega_{30}$             \\ \hline
 &  &  & \textbf{I} & 4.90~[GHz]  & 4.47~[GHz]            & 4.06~[GHz]     & 13.43~[GHz] \\ \hline
 &  &  & \textbf{II} & 3.53~[GHz]  & 5.22~[GHz]            & 7.04~[GHz]     & 15.8~[GHz]   \\ \hline
 &  &  &  \textbf{III}& 2.09~[GHz]  & 2.44~[GHz]            & 0.61~[GHz]     & 5.15~[GHz]  \\ \hline
\end{tabular}}
\caption{Optimal circuit topology, circuital parameters, and relevant transition frequencies for a single-loop circuit containing two (\textbf{I}) and three (\textbf{II}) links, and for a one-dimensional array of two sites (\textbf{III}) that describes a ladder three-level system.}
\label{Table1}
\end{table}
\begin{table}[t]
\centering
\scalebox{0.9}{
\begin{tabular}{|c|l|l|l|l|l|l|l|}
\hline
\multicolumn{1}{|l|}{\begin{tabular}[c]{@{}l@{}}$(\mathcal{T}^{\star},\mathcal{P}^{\star})$\end{tabular}} & $\alpha$ & $\beta$ & $\gamma$ & $C~\rm{[pF]}$ & $L~\rm{[nH]}$ & $C_{J}~\rm{[pF]}$ & $E_{J}/\hbar~\rm{[GHz]}$ \\ \hline
\multicolumn{8}{|c|}{\textbf{IV}}                                             \\ \hline
0 & 0 & 1 & 0 & 0             & 150            & 0            & 0              \\ \hline
1 & 0 & 0 & 0 & 0             & 0            & 0& 0              \\ \hline
2 & 0 & 0 & 0 &  0 &0            & 0            & 0  \\ \hline
3 & 0 & 0 & 1 & 0             & 0            & 0.028           & 64.21              \\ \hline
\multicolumn{8}{|c|}{\textbf{V}}                                          \\ \hline
0 & 0 & 0 & 1 & 0  & 0 & 0.015 & 100 \\ \hline
1 & 0 & 1 & 0 & 0  & 19.84 & 0 & 0 \\ \hline
2 & 0 & 0 & 0 & 0  & 0 & 0 & 0  \\ \hline
3 & 1 & 0 & 0 & 0.045  & 0 & 0 & 0  \\ \hline
\multicolumn{8}{|c|}{\textbf{VI}}                                             \\ \hline
\multicolumn{8}{|c|}{\textbf{Box 1}}                                             \\ \hline
0 & 0 & 0 & 1 & 0             & 0            & 0.0021            & 100             \\ \hline
1 & 0 & 1 & 0 & 0             & 0.24            & 0& 0              \\ \hline
2 & 0 & 0 & 0 & 0 &0       & 0            & 0  \\ \hline
3 & 1 & 0 & 0 & 0.00015             & 0            & 0          &0             \\ \hline
\multicolumn{8}{|c|}{\textbf{Box 2}}                                            \\ \hline
0 & 0 & 0 & 1 & 0             & 0            & 0.356  & 34.28 \\ \hline
1 & 0 & 1 & 0 & 0             & 113 & 0            & 0              \\ \hline
2 & 0 & 0 & 0 & 0             & 0& 0            & 0              \\ \hline
3 & 1 & 0 & 0 & 0.00015             & 0            & 0            & 0              \\ \hline
\multicolumn{8}{|c|}{\textbf{Coupling}}                                           \\ \hline
0 & 1 & 0 & 0 & 0.173  & 0 & 0 & 0              \\ \hline
\multicolumn{8}{|c|}{\textbf{System parameters }}                                           \\ \hline
 &  &  &  & $\omega_{10}$  & $\omega_{21}$           & $\omega_{20}$           & $\omega_{30}$             \\ \hline
 &  &  & \bf{IV}& 0~[GHz]  & 18.27~[GHz]            & 18.27~[GHz]      & 18.27~[GHz] \\ \hline
 &  &  &\bf{V} & 0~[GHz]  &5.07~[GHz]          & 5.07~[GHz]     & 5.07~[GHz]   \\ \hline
 &  &  & \bf{VI}& 0~[GHz]  &13.3~[GHz]            & 13.3~[GHz]     & 15.9~[GHz] \\ \hline
\end{tabular}}
\caption{Optimal circuit topology, circuital parameters, and relevant transition frequencies for a single-loop circuit containing two (\textbf{IV}) and three (\textbf{V}) links, and for a one-dimensional array of two sites (\textbf{VI}) that describe a lambda system.}
\label{Table2}
\end{table}
Table~\ref{Table1} shows that the optimal configuration for a $\Xi$ three-level system consisting of two links (\textbf{I}) corresponds to a pair of coupled Josephson junctions threaded by an external magnetic flux. Notice that this configuration is similar to the split-transmon circuit or the Rf-SQUID~\cite{Orlando1999,Mooij1999} architecture corresponding to archetypal charge and flux qubit, respectively. In these systems, the low-lying energy spectrum has large anharmonicity that guarantees to perform logic operations on them. Furthermore, their charge operators follow the same selection rules as a $\Xi$ three-level system. Moreover, the minimization algorithm extends the previous result for a single-loop device consisting of three links (\textbf{II}); the optimal circuit corresponds to a pair of Josephson junctions coupled to a capacitor. In this architecture, the additional capacitor changes the ratio between the total Josephson energy with the respective charge energy, similar to the transmon circuit. In such a device, the charge operator also demonstrates the same selection rules as the $\Xi$ three-level system. Finally, for the extended circuit (\textbf{III}), the optimal architecture corresponds to the capacitive coupling between a single-Josephson junction with a circuit consisting of two inductors connected to a capacitor and an additional Josephson junction.\par

On the other hand, for the $\Lambda$ three-level system that consists of a single-loop circuit with two links (\textbf{IV}), the optimal architecture is like a Fluxonium configuration~\cite{Nguyen2019} formed by a Josephson junction coupled to an inductor. Notice that this configuration behaves as a $\Lambda$ three-level system for an external magnetic flux fixed at $\varphi_{x}/\pi=\varphi_{0}/2$~\cite{PhysRevLett.95.087001}. Like the $\Xi$ three-level system, we obtain that the optimal circuit configuration having three links (\textbf{V}) is the extension of the previous one, where we couple an additional capacitor to the Josephson junction and the inductor, respectively. Finally, the optimal configuration for an extended circuit (\textbf{VI}) corresponds to a pair of Fluxonium coupled capacitively with each other. As expected, our circuital parameters are consistent with the reported values in the previous cQED implementations.\par

To compare the performance of each optimal circuit architecture, we will calculate at zero $\varphi_{x}=0$ external flux, the energy differences and the relative anharmonicity between the low-lying energy spectrum defined as $\mathcal{A}_{ij,kl}=(\omega_{ij}-\omega_{kl})/(\omega_{ij}+\omega_{kl})$, with $\omega_{ij}$ the transition frequency between the $i$th and $j$th energy states, respectively. Let us start with the $\Xi$ three-level system 
\begin{subequations}
	\begin{eqnarray}
		\mathcal{A}_{21,10}^{\Xi}&=&\{-0.046,0.193,0.077\},\\
		\mathcal{A}_{32,21}^{\Xi}&=&\{-0.048,0.148,-0.60\}.
	\end{eqnarray}
\end{subequations}
For circuit (\textbf{I}), we obtain similar values for the anharmonicity. These results confirm that the device behaves as a split transmon circuit since the first three lowest energy transitions are identical. Consequently, we can model the device as a weakly anharmonic oscillator. Moreover, we obtain significantly larger anharmonicity for the second circuit (\textbf{II}) than the device (\textbf{I}), respectively. Each device's successive energy transitions are different, making it unsuitable for the required $\Xi$ three-level system. Finally, we achieve better performance for the extended multi-loop circuit (\textbf{III}) than the other two configurations. We obtain a low anharmonicity between the first and the second excited state $\mathcal{A}_{21,10}^{\Xi}=0.077$, whereas the third energy level off-resonance from the two different low-lying energy levels $\mathcal{A}_{32,21}^{\Xi}=-0.60$. Thus, the extended multi-loop circuit attains the required spectral configuration for the $\Xi$ three-level system.\par

We extend a similar analysis for the $\Lambda$ configuration. However, instead of calculating the anharmonicity implying the first energy transition, we consider the energy differences mediated by the second excited state as $\omega_{20}$ and $\omega_{21}$. Moreover, as pointed out in Table~\ref{Table2}, we obtain that for the lambda circuits containing two (\textbf{IV}) and three links (\textbf{V}). The relative anharmonicity is zero, which corresponds to a quantum system with a degenerate ground state. Notice that this degeneracy also appears on the second and third excited states where the relative anharmonicity is zero. That condition makes it unsuitable for a lambda configuration since it is very likely that these degenerate energy states will participate in the system dynamics. For the extended multi-loop $\Lambda$ configuration, we obtain that the relative anharmonicity between the transitions $\omega_{20}$ and $\omega_{21}$ is also zero. Nevertheless, the one-dimensional array breaks the degeneracy between the second and the third excited state, obtaining a relative anharmonicity $\mathcal{A}_{30,20}^{\Lambda}=\mathcal{A}_{30,21}^{\Lambda}=0.1$ that permit to avoid the dynamics involving higher excited energy levels.\par

We conclude that, in multi-loop circuits (\textbf{III} and \textbf{VI}), our algorithm finds architectures with better performance in comparison with the single-loop circuit having two and three links each. We attribute such improvement to the presence of redundancy that involves having several degrees of freedom on the device. This redundancy creates manifolds whose energy spectrum is robust against fluctuations in the physical parameters leading to noise protection~\cite {Gyenis2021}. 

\section{Dependence on the external magnetic flux}\label{V}
The next step in the characterization of the optimal device obtained with our algorithm relies on its response (tunability) of the energy spectrum and transition matrix elements of the charge operator $\hat{N}$ when an external phase thread either the single-loop or the extended one-dimensional array. Here, we consider the number operator at the edges of both configurations (the single-loop and the extended one-dimensional system). Moreover, we use a linear modulation for the external phase of the form $\phi_{x}/\phi_0=\varphi_x=(-2\pi,2\pi)$.\par
\begin{figure}[!t]
\centering
\includegraphics[width=1\linewidth]{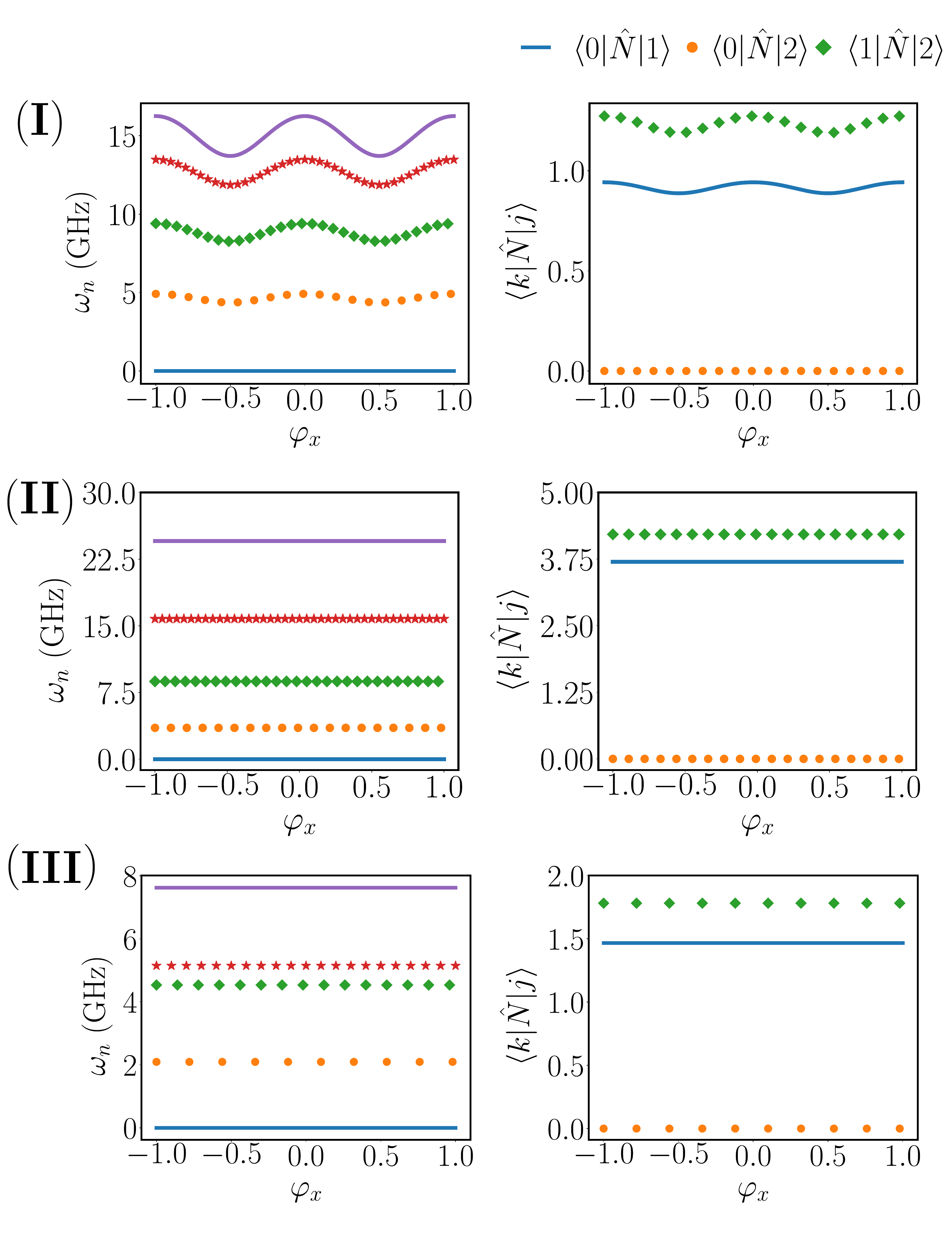}
\caption{(Color online) Energy spectrum and relevant transition matrix elements $\langle0|\hat{N}|1\rangle$ (yellow), $\langle1|\hat{N}|2\rangle$ (blue) and $\langle0|\hat{N}|2\rangle$ (green) for the node charge operator $\hat{N}$ as a function of the frustration $\varphi_{x}$, and for the optimal circuits describing a ladder three-level system for architectures containing two (\textbf{I}), three (\textbf{II}),  and the one-dimensional array (\textbf{III}) with two sites. The circuital parameters are taken from Table~\ref{Table1}. We observe a trade-off between achieving the required energy configuration and matching the needed selection rules in single-loop circuits with two and three branches, respectively. For the multi-loop system, we do not observe this trade-off. Consequently, the multi-loop architecture matches both requirements in the energy spectrum and selection rules for all control parameter values.}
\label{fig:fig4}
\end{figure}
Figure \ref{fig:fig4} depicts the low-lying energy spectrum and the transition matrix elements [$\langle0|\hat{N}|1\rangle$, $\langle1|\hat{N}|2\rangle$, $\langle0|\hat{N}|2\rangle$] for the single-loop circuit containing two links (\textbf{I}), three links (\textbf{II}), and the multi-loop linear array (\textbf{III}), respectively. For the device (\textbf{I}), we observe that the studied quantities changes with the external flux exhibiting an oscillatory behavior. Nevertheless, the changes are sufficiently small not to appreciate avoided energy spectrum. Moreover, as pointed out in the previous section, we also see low anharmonicity between the successive energy levels. We also see that the transition matrix elements satisfy the requirement for a $\Xi$ three-level system where the transition $\ket{0}\leftrightarrow\ket{2}$ is highly suppressed since its matrix element is always zero. Also, the transition matrix elements $\langle0|\hat{N}|1\rangle$, and  $\langle1|\hat{N}|2\rangle$ slightly oscillate, guaranteeing transitions between consecutive energy levels. 

We see the absence of such oscillations in the low-lying energy spectrum and transition matrix elements for the three-links circuit (\textbf{II}) and the multi-loop device (\textbf{III}). In other words, the studied quantities become insensitive to fluctuations in the control parameter, as depicted in Fig. \ref{fig:fig4}~(\textbf{II}) and Fig. \ref{fig:fig4}~(\textbf{III}), respectively. These architectures also satisfy the conditions to behave as a $\Xi$ three-level system where the matrix element $\langle0|\hat{N}|2\rangle$ is suppressed.\par
\begin{figure}[!t]
\centering
\includegraphics[width=1\linewidth]{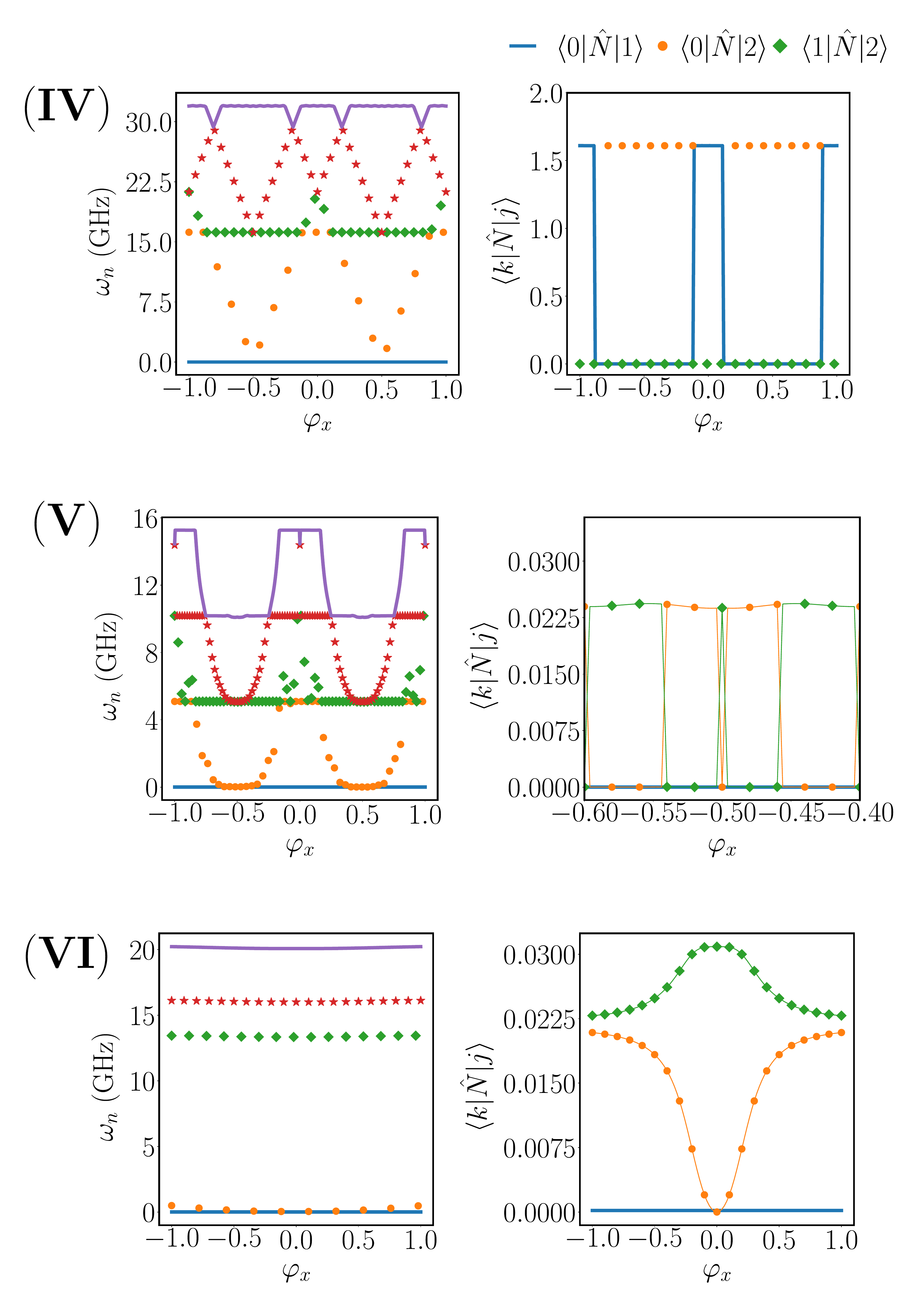}
\caption{(Color online) Energy transition and relevant transition matrix elements $\langle0|\hat{N}|1\rangle$ (yellow), $\langle2|\hat{N}|1\rangle$ (blue), and $\langle2|\hat{N}|0\rangle$ (green) for charge operator $\hat{N}$ as a function of the external magnetic flux $\varphi_{x}$, and for the optimal circuits describing a lambda three-level system for architectures containing two (\textbf{IV}), three (\textbf{V}), and the one-dimensional array (\textbf{VI}) with two sites. The circuital parameters are taken from Table~\ref{Table2}. We see the competition between accomplishing the demanding energy configuration while satisfying the selection rules in circuits containing two and three branches. For the extended architecture, we do not see that competition. Instead, the device achieves the desired energy landscape with the proper selection rules.}
\label{fig:fig5}
\end{figure}
We extend the same analysis to the optimal architecture for the $\Lambda$ system. Fig.~\ref{fig:fig5} shows the low-lying energy spectrum and the transition matrix elements ($\langle0|\hat{N}|1\rangle$, $\langle1|\hat{N}|2\rangle$,$\langle0|\hat{N}|2\rangle$) for a single-loop having two and three links, as well as the multi-loop $\Lambda$ configuration. 

For both single-loop circuits (\textbf{VI}) and (\textbf{V}), we observe that the low-lying energy spectrum and the transition matrix elements present \textit{sweet spots} around $\varphi_{x}=\pm 0.5$. At this value, the first two energy levels mutually approach ($\omega_{10}\sim0$), forming quasi-metastable energy levels, as depicted in Fig.~\ref{fig:fig5}. At these values, the relevant transition matrix elements satisfy the desired criteria at these values where the transition matrix element $\langle0|\hat{N}|1\rangle$ vanishes, as shown in Fig.~\ref{fig:fig5}. 

Finally, we consider the multi-loop with the parameters from Table~\ref{Table2}. For this architecture, the energy spectrum becomes quasi-insensitive to the frustration parameter. Even though the energy spectrum of the artificial atom changes with $\varphi_{x}$, its energy differences remain constants (see panel (\textbf{VI}) from Fig.~\ref{fig:fig5}). Moreover, the relevant transition matrix element of both charge and flux operator vanishes at $\varphi_{x}=0$. For another value of the frustration parameter, the system does not behave as a $\Lambda$ three-level system; for other values of the external magnetic flux, the transition matrix elements satisfy $\hat{N}_{20}\neq0$, $\hat{N}_{21}\neq0$ and $\hat{N}_{10}=0$.\par

We conclude that, for circuits containing a single quantum degree of freedom (\textbf{I} and \textbf{IV}), it is hard for the algorithm to find a suitable architecture in which both the atomic energy spectrum and its correspondent selection rules are attained. We can only obtain sweet spots where the system mimics the atomic system. In contrast to circuits with fewer degrees of freedom, those with more offer a wealth of potential parameter configurations. This can manifest as spectral manifolds or protected subspaces~\cite{Gyenis2021,Douzot}, enabling the design of circuits with specific spectral characteristics and selection rules. Additionally, the added redundancy in the degrees of freedom makes these circuits more resilient, allowing for 'sweet regions' of operation in which the circuit behaves in a way that is similar to an atomic system.
\begin{figure}[!t]
\centering
\includegraphics[width=1\linewidth]{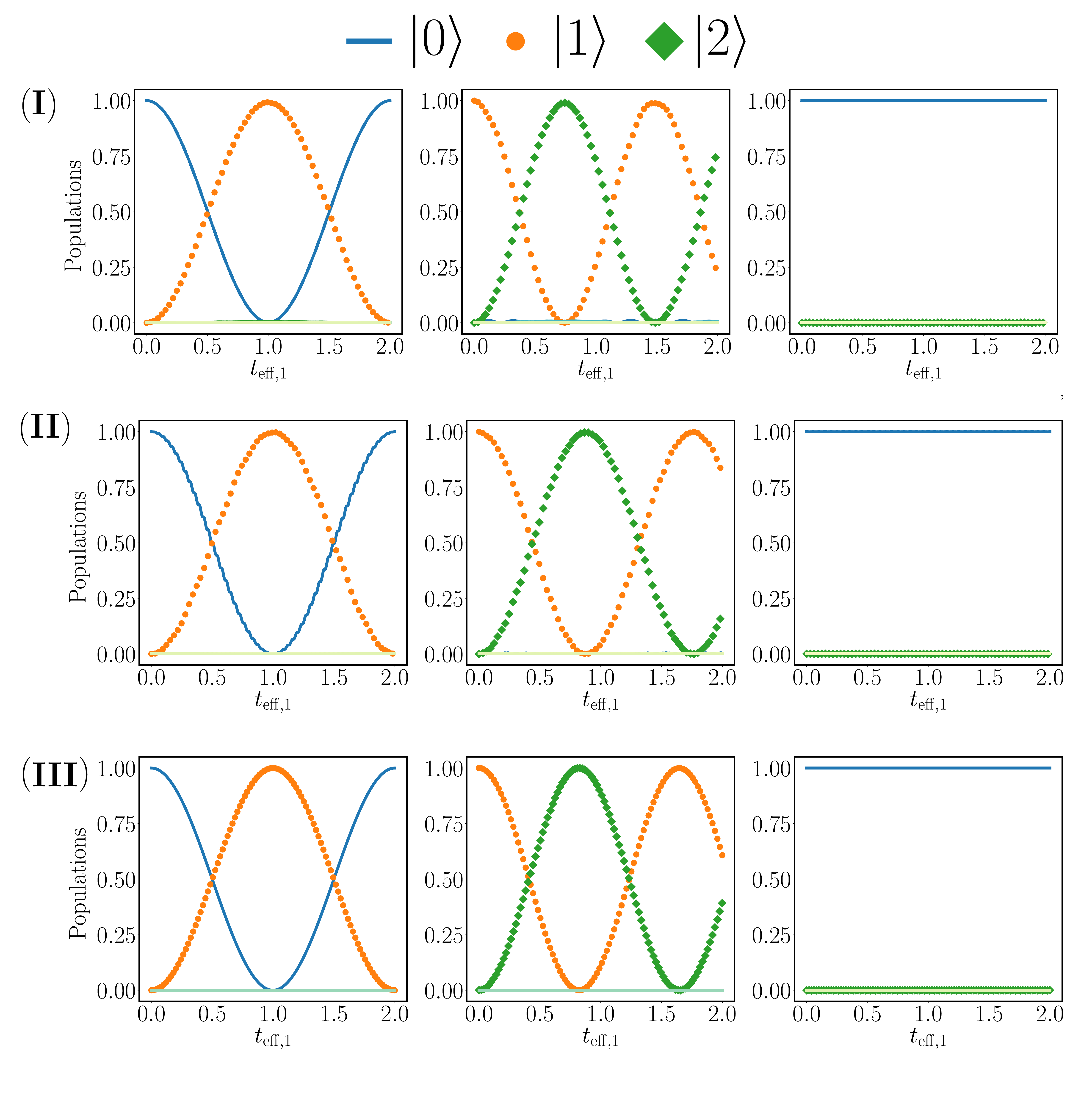}
\caption{(Color online) Population evolution of the low-lying energy states $\ket{\Psi_{j}}$ of the ladder configuration numerically calculated with the Hamiltonian in Eq.~(\ref{driving}) for the circuit containing (I) two links, (II) three links, and (III) the one-dimensional array has two sites. The columns represent the evolution considering the driving resonance with the energy transitions $\nu=\omega_{10}$, $\nu=\omega_{21}$, and $\nu=\omega_{20}$, respectively. The system parameters of the circuit are the same as in Table~\ref{Table1} for external value $\varphi_{x}=0$.}
\label{fig:fig6}
\end{figure}
\begin{figure}[t]
\centering
\includegraphics[width=1\linewidth]{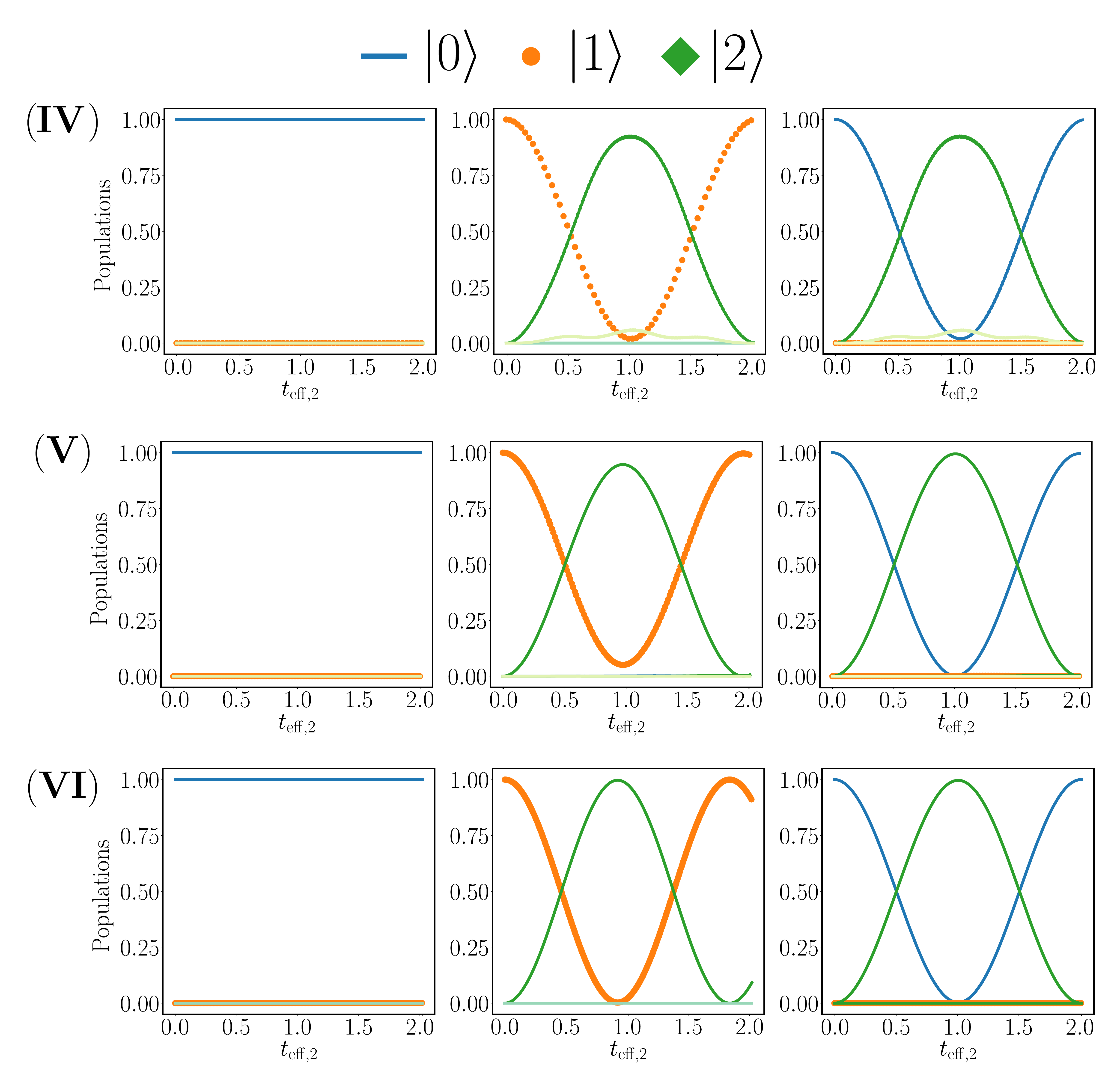}
\caption{(Color online) Population evolution of the low-lying energy states $\ket{\Psi_{j}}$ of the $\Lambda$ configuration numerically calculated with the Hamiltonian in Eq.~(\ref{driving}) for the circuit containing (I) two links, (II) three links, and (III) the multi-loop circuit has two sites. The columns here represent the evolution considering the driving resonance with the energy transitions $\nu=\omega_{10}$, $\nu=\omega_{21}$, and $\nu=\omega_{20}$, respectively. The system parameters of the circuit are the same as Table~\ref{Table1}.}
\label{fig:fig7}
\end{figure}

\section{Dynamical properties}\label{VI}
We study the dynamics of the optimal circuits under the action of a driving tuned to the relevant transition acting on the charge operator $\hat{N}$ described by the Hamiltonian of the system reads 
\begin{eqnarray}
\label{driving}
\bar{\mathcal{H}} &=& \mathcal{H}(t) + \Omega\cos(\nu t)\hat{N},
\end{eqnarray}
where $\mathcal{H}$ is the Hamiltonian of the optimal single-loop and multi-loop configuration obtained in Table~\ref{Table1} and Table~\ref{Table2}, moreover, $\Omega$ is the driving strength, and $\nu$ corresponds to the driving frequency, which we have selected to be $\nu=\omega_{10}$, $\nu=\omega_{21}$ or $\nu=\omega_{20}$. We calculate the probability evolution of the eigenstates of $\mathcal{H}$ by initializing the system in its ground state $\ket{0}$ for the driving frequencies $\nu=\omega_{10}$ and $\nu=\omega_{21}$, whereas we initialize the system and in $\ket{1}$ for $\nu=\omega_{20}$, respectively.\par
\begin{figure}[b]
\centering
\includegraphics[width=1\linewidth]{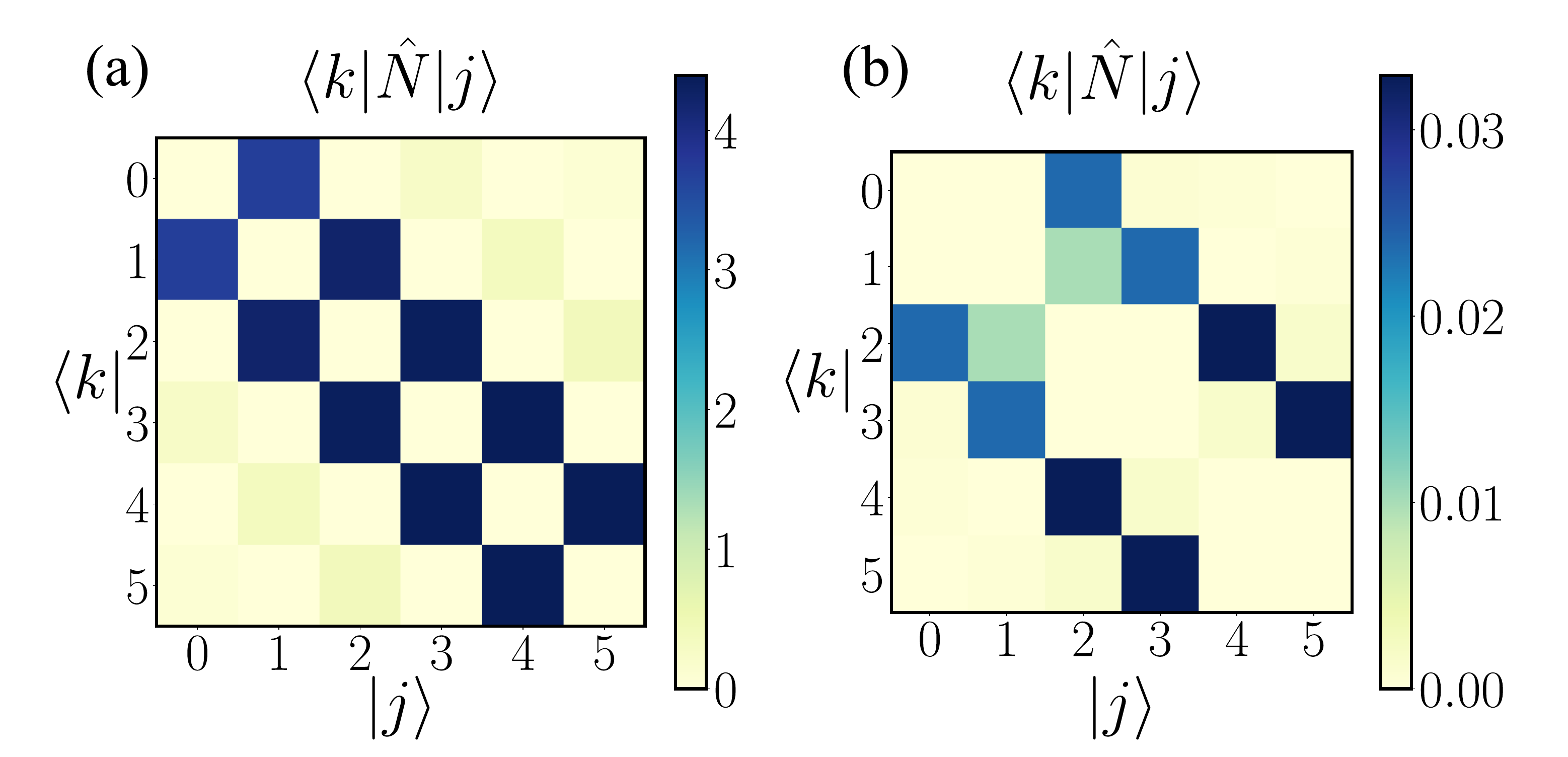}
\caption{(Color online) Matrix representation for the left flux and charge operator of the multi-loop system containing four single-loop circuits with the parameters obtained in Table~\ref{Table1} and Table~\ref{Table2} for ladder (a) and lambda (b) system at frustration parameter $\varphi_{x}=1$.}
\label{fig:fig10}
\end{figure}
Figure~\ref{fig:fig6} shows the population evolution for the circuits (\textbf{I}), (\textbf{II}) and (\textbf{III}) biasing $\varphi_{x}=0$, in units of the dimensionless time $t_{\rm{eff},1}=\pi/(\Omega\hat{N}_{10,\ell})$, where $N_{10,\ell}$ corresponds to the transition matrix element for the node charge operator. 

From a dynamical point of view, we see that every circuit configuration found by our algorithm behaves as $\Xi$ three-level system; depending on the driving frequency, the device access to adjacent energy states $\{\ket{0},\ket{1}\}$, and $\{\ket{1},\ket{2}\}$ without inducing transitions involving additional higher excited energy levels. In fact, for panel (I) regarding the single-mode circuit, the population trapped in the second excited state is negligible in comparison with the population in the first excited state, observing an almost-complete Rabi oscillation between the states $\{\ket{0},\ket{1}\}$. For multi-loop circuits (panel (III)), we observe that the higher energy transitions are suppressed, obtaining Rabi oscillations between the adjacent levels. We understand this by looking at Fig.~\ref{fig:fig10} where we have plotted the transition matrix element for the phase and charge operator for the circuit (III), observing that the dominant transition matrix element corresponds to the adjacent states $\ket{j}$ and $\ket{j+1}$. In this scenario, transition involving more than one excitation is suppressed because it corresponds to a slow process that will not occur within the addressing time scale provided by the driving. Another reason for obtaining the system dynamics in a reduced Hilbert space relies on the anharmonicity between the excited energy levels.\par
We extend the same analysis to the $\Lambda$ configuration showing the system dynamics of the circuits (\textbf{IV}), (\textbf{V}) and (\textbf{VI}) in Fig.~\ref{fig:fig7}. We observe a poor performance for the single-mode circuits from a dynamic point of view. Even though it is possible to suppress the $\ket{0}\leftrightarrow\ket{1}$ transition (there is no evolution of the probability), we observe population leakage and trapping in the higher excited energy states. Consequently, we see an imperfect Rabi oscillation between the resonance eigenstates. We appreciate the absence of trapping and leakage in the higher energy states for the multi-loop circuit as depicted in Fig.~\ref{fig:fig10}{(\textbf{VI})}. Thus, the redundancy in terms of the degree of freedom allows us to design architectures achieving both the desired energy spectrum and the dynamic properties of an atomic system.
\begin{figure}[t]
\centering
\includegraphics[width=1\linewidth]{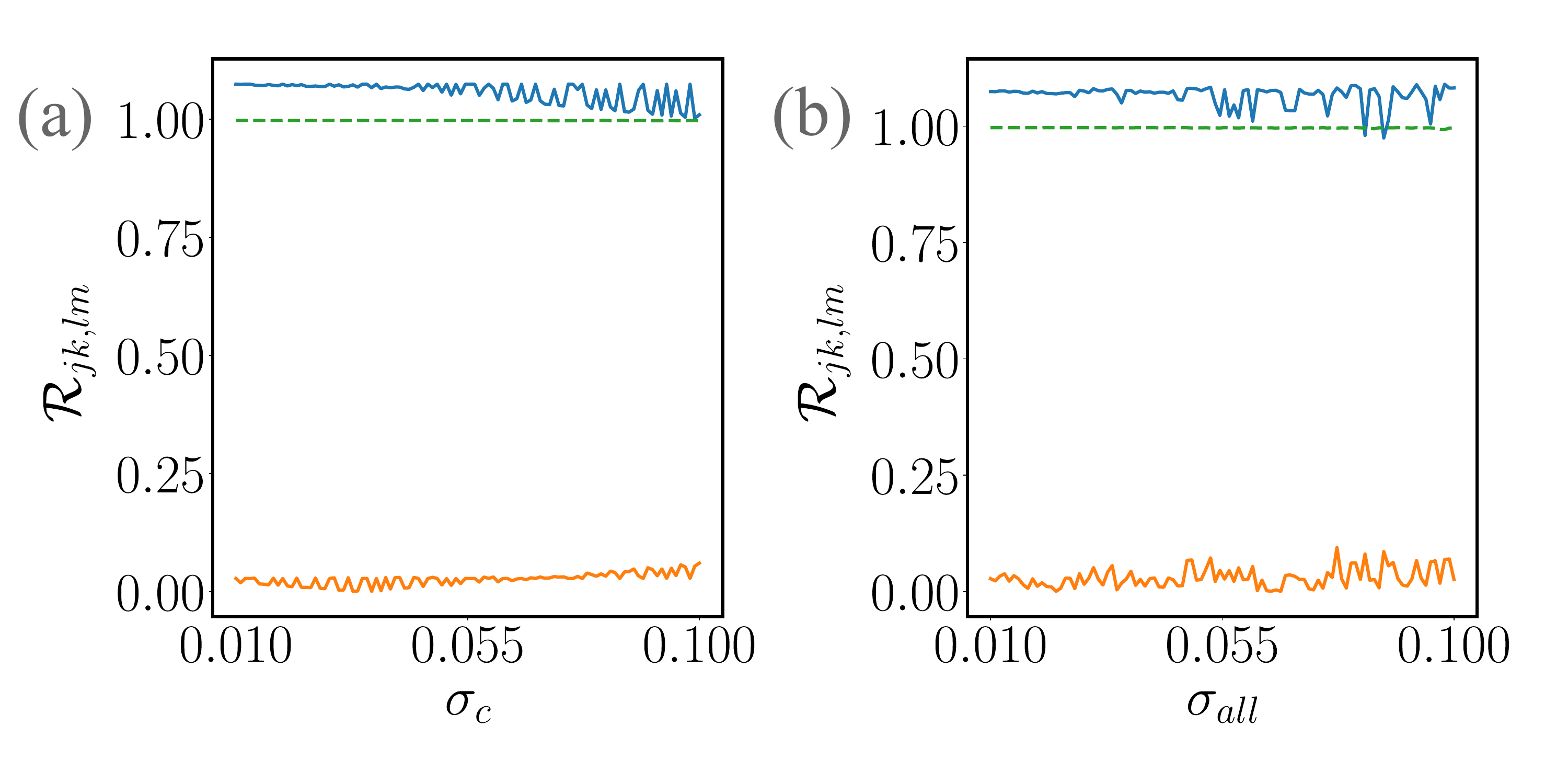}
\caption{(Color online) Relative energy transitions $\mathcal{R}_{ij,lk}$ as a function of (a) the capacitance error $\sigma_{c}$ and (b) the error in all the circuital parameters $\sigma_\text{all}$ for the optimal multi-loop circuits given in Table~\ref{Table1} and Table~\ref{Table2}. Solid blue and orange lines correspond to the ratios $\mathcal{R}_{21,10}$ and $\mathcal{R}_{32,21}$, respectively. We expect that a ladder three-level system $\Xi$ has good performance when $\mathcal{R}_{21,10}\approx1$ and $\mathcal{R}_{32,21}\neq1$ since these conditions represent that the first two energy transitions are identical and the third energy level is far off-resonance. The green dashed line is the ratio $\mathcal{R}_{21,20}$ describing the three-level lambda $\Lambda$. We achieve two metastable ground states when $\mathcal{R}_{21,20}\approx1$.}
\label{fig:fig8}
\end{figure}

\section{Resilience against circuit parameter fluctuations}\label{VII}
We evaluate the robustness of our protocol against parameter fluctuation on their circuital parameters. We will focus on the multi-loop cases, which have demonstrated better performances than the single-mode configurations. We start with controlled errors in one of the capacitors and then extend to all the parameters. Finally, we analyze the performance of the multi-loop architecture in which all parameters fluctuate. 

For the controlled errors, we assume random deviations around $\pm10\%$ of the optimal parameters obtained in Table~\ref{Table1}, and Table~\ref{Table2} for the multi-loop case, respectively. For all the parameters fluctuating, we consider a normal distribution of the optimal parameters with a deviation of $\pm5\%$ around the optimal values. We evaluate the performance of the multi-loop circuit in terms of its energy transition through the following ratio.
\begin{eqnarray}
\mathcal{R}_{jk,lm} = \frac{\omega_{j}-\omega_{k}}{\omega_{l}-\omega_{m}}\geq0;\quad (j>k, l>m).
\end{eqnarray}
\begin{figure*}[t]
	\centering
	\includegraphics[width=1\linewidth]{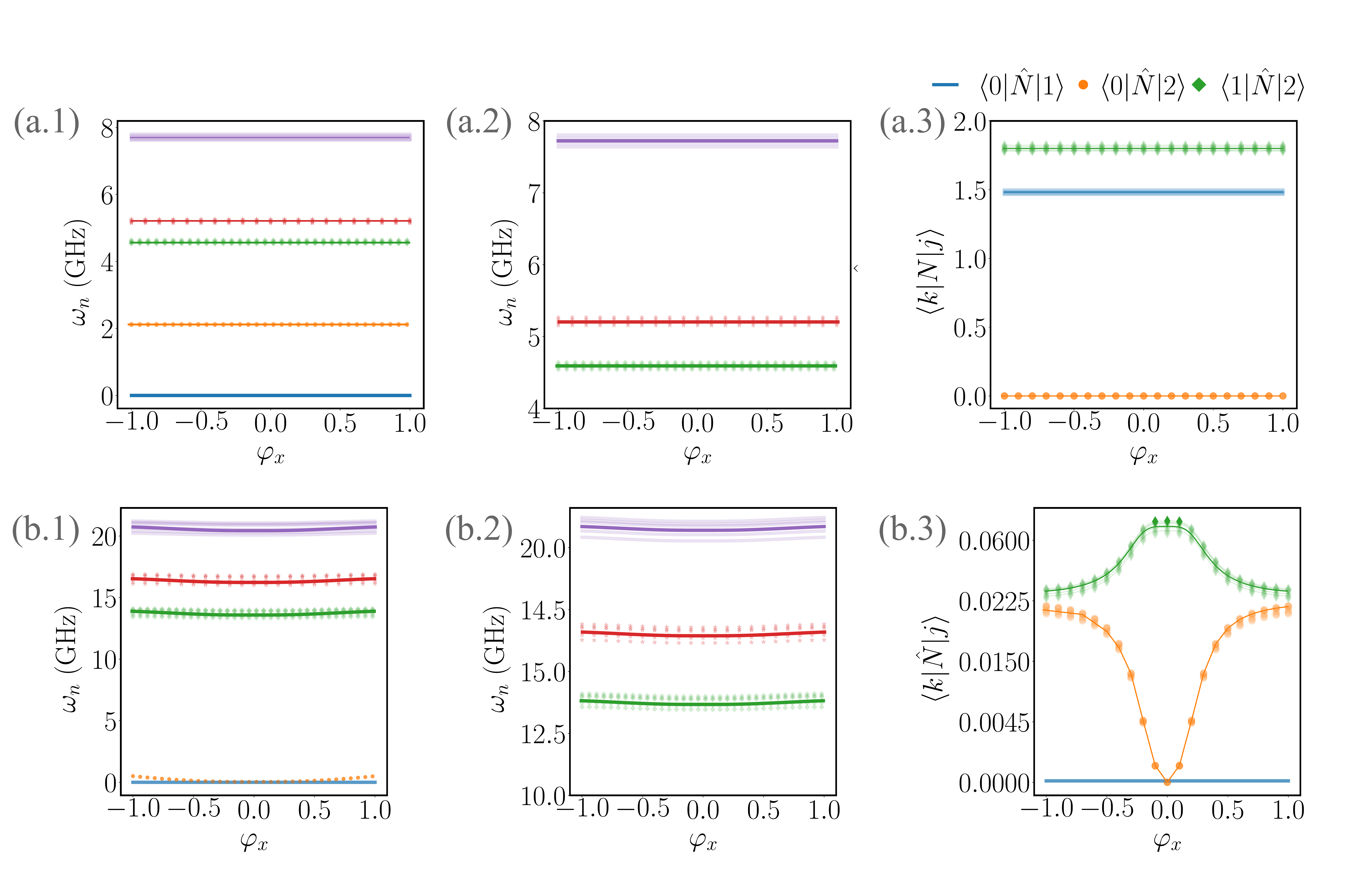}
	\caption{(Color online) (a.1) Energy spectrum for the optimal multi-loop $\Xi$ three-level system ($\bf{III}$) and (b.1) the multi-loop $\Lambda$ three-level system ($\bf{VI}$) as a function of the external phase $\varphi_{x}$. We have considered  100 different random circuit parameters following a normal distribution with average as the parameters obtained in Table 1 and Table 2 and standard deviation around 5$\%$. (a.2) and (b.2) show an inline of the energy spectrum that demonstrates no changes in the relative anharmonicity even though the spectrum fluctuates for different parameters. Finally, (a.3) and (b.3) depict the relevant transition matrix elements as a function of the external phase $\varphi_{x}$ for the aforementioned random circuit parameters. In the figures, the solid color line corresponds to the average value of the energy spectrum and transition matrix elements, whereas the the less opaque curves corresponds to fluctuation around the average value, respectively.}
	\label{fig:fig9}
\end{figure*}
Where $\omega_{\ell}$ is the $\ell$th transition of the multi-loop circuit. For an ideal $\Xi$ three-level system, we expect $\mathcal{R}_{21,10}\approx1$ meaning that the energy transitions $\omega_{21}$ and $\omega_{10}$ have identical transition frequencies. Moreover, we also demand that $\mathcal{R}_{32,21}\neq 1$ that characterize the discrepancy between the energy transitions $\omega_{32}$ and $\omega_{21}$ such that the former be not accessible during the dynamics. Notice that $\mathcal{R}_{32,21}$ could be larger or smaller than one depending on which energy transition has a larger transition frequency. Contrary, for an ideal $\Lambda$ three-level system, we demand that $\mathcal{R}_{21,20}\approx1$, which means that the energy transitions $\omega_{21}$ and $\omega_{20}$ are similar.\par

Figure.~\ref{fig:fig8}{(a)} shows the set of ratios $\mathcal{R}_{21,10}$, $\mathcal{R}_{32,21}$, and $\mathcal{R}_{21,20}$ as a function of the capacitance error $\sigma_{c}$, whereas Fig.~\ref{fig:fig8}{(b)} show them as a function of the total error $\sigma_{c}$, respectively. The figures show that both multi-loop three-level systems are resilient against these controlled parameter errors. The $\Xi$ three-level system depicts that the ratio $\mathcal{R}_{21,10}$ has a stable value for small controlled errors ($\sigma_{c}=0.055$). In contrast, it shows larger fluctuations around for controlled errors which are larger than this value. However, even though these fluctuations are noticeable, they converge to $\mathcal{R}_{21,10}=1$, which is the condition needed for a good $\Xi$ three-level system. For the ratio $\mathcal{R}_{32,21}$, we observe fluctuations around zero, ensuring that the adjacent energy transition and its respective matrix elements of the three-level system keeps the same structure with similar transition frequencies also being far off-resonance with higher energy transitions.\par

We extend the discussion for the multi-loop $\Lambda$ three-level system, also showed in Fig.~\ref{fig:fig8}{(a)-(b)}. The figure shows that this architecture is more resilient against controlled errors since the ratio $\mathcal{R}_{21,20}$ keeps a constant value around one for all the analyzed ranges. Then, our automated circuit quantization algorithm gives us a set of three-level systems with optimal circuit architecture resilient against controlled fluctuation on the circuital parameters.

We provide additional evidence of resilience against fluctuations by calculating the energy spectrum and the transition matrix element of both optimal circuits (\textbf{III}) and (\textbf{VI}). We consider a data set of 100 samples whose parameters follow a normal distribution averaged with the parameters obtained in Table~\ref{Table1}, and Table~\ref{Table2} with deviations around $5\%$. We summarize our findings in Fig.~\ref{fig:fig9}, where we have plotted the energy spectrum and the transition matrix elements for all those configurations.\par

For the multi-loop $\Xi$ three-level system, Fig.~\ref{fig:fig9}{(a.1)} shows that the first two energy levels are more resilient against fluctuations since these energy transitions are almost constant for all the circuits of the dataset. The effect of the deviations appears from the second excited state, where the energies oscillate around an equilibrium configuration consisting of the average of these transitions without changing the anharmonicity of these states as depicted in Fig.~\ref{fig:fig9}{(a.2)} where we show an inline of Fig.~\ref{fig:fig9}{(a.1)} with the highest energy levels that confirm our statement. For the transition matrix elements, we appreciate that it follows the exact behavior of the low-lying energy levels; the values fluctuate around the average, but still, the circuit maintains the selection rules imposed by the algorithm, see Fig.~\ref{fig:fig9}{(a.3)}.\par

For the multi-loop $\Lambda$ three-level system, we observe a similar tendency to the ladder three-level system. The first two energy levels are resilient against fluctuations because they are constant for all the circuits of the dataset for all the values of the control parameter. Only for the highest energy levels can we observe the effects of the deviations. See Fig.~\ref{fig:fig9}{(b.1)} and the inline depicted in Fig.~\ref{fig:fig9}{(b.2)}. Finally, for the transition matrix elements, we appreciate that it follows the exact behavior of the low-lying energy levels.\par

In conclusion, the multi-loop $\Lambda$ three-level system performs better against controlled errors since it does not exhibit appreciable changes in the figure-of-merit used to quantify its resilience. On the contrary, we observe larger fluctuations for the $\Xi$ three-level system, but they converge to the expected value. 

For the random sampling of the circuit parameters, we observe identical performance for both multi-mode circuits; the fluctuations in the energy levels start to appear in the second excited state of both configurations. Nevertheless, changes on the spectrum do not modify the anharmonicity between them since they all happen in the same proportion for deviations around 5$\%$. We observe similar behavior for the transition matrix elements, where the presence of the fluctuations does not alter the selection rules of the multi-loop configuration. Thus, designing architectures with engineered energy levels and selection rules within consistent cQED parameters with bounded fabrication errors is possible.
\section*{Conclusions}
In this work, we have used genetic algorithms to design superconducting quantum circuits with atomic energy spectra and selection rules mimicking ladder and lambda three-level systems. The algorithm starts by developing an automatized circuit quantization subroutine to calculate the quantum Hamiltonian of a multi-loop system composed of randomly selected circuit elements. Our approach can choose an adequate quantization basis for each degree of freedom constituting the multi-loop system relying upon the circuit configuration. Then, the genetic algorithm adapts the circuit topology and the circuital parameters so that its energy spectrum and transition matrix elements behave as desired, i.e. as a specific three-level system.

We have found that in single-loop configurations, there exists a tradeoff between attaining the spectral requirement of the circuit and meeting the desirable conditions for the transition matrix elements. The conclusion is that the system has too few free parameters to fulfill all conditions simultaneously. We circumvent this problem by using a multi-loop system comprising two single-loop systems. With this setup, we have obtained that it is possible to satisfy all the conditions for a wide range of the control parameter named sweet regions. Additionally, we demonstrated that our multi-loop circuits are robust against random fluctuations in their circuit parameters, such as fabrication errors, making them suitable for use in large-scale setups as modular components with specific symmetries.
\section*{ACKNOWLEDGMENT}

F.A.C.L. acknowledges F. Motzoi for helpful discussions. The authors acknowledge support from the German Ministry for Education and Research, under QSolid Grant No. 13N16149, the Chilean Government \textit{Financiamiento Basal para Centros Científicos y Tecnológicos de Excelencia} (Grant No. FB0807), USA2055\_DICYT, Universidad de Santiago de Chile. We acknowledge support from the HORIZON-CL4-2022-QUANTUM01-SGA project 101113946 OpenSuperQ-Plus100 of the EU Flagship on Quantum Technologies, the Spanish \textit{Ramón y Cajal} Grant RYC-2020-030503-I, and the “Generación de Conocimiento” project Grant No. PID2021-125823NA-I00 funded by MICIU/AEI/10.13039/501100011033, by “ERDF Invest in your Future,” and by FEDER EU. We also acknowledge support from the Basque Government through Grants No. IT1470-22, the Elkartek project KUBIT KK-2024/00105, and from the IKUR Strategy under the collaboration agreement between Ikerbasque Foundation and BCAM on behalf of the Department of Education of the Basque Government. This work has also been partially supported by the Ministry for Digital Transformation and the Civil Service of the Spanish Government through the QUANTUM ENIA project call – Quantum Spain project, and by the European Union through the Recovery, Transformation and Resilience Plan – NextGenerationEU within the framework of the Digital Spain 2026 Agenda.

\section*{COMPETING INTERESTS}
The authors declare no competing interests.
\section*{DATA AVAILABILITY}
The data that support the findings of this study are available on the Zenovo repository~\cite{Data}.
\appendix
\section{Eliminating passive nodes}\label{A1}
\begin{figure}[t]
\centering
\includegraphics[width=.7\linewidth]{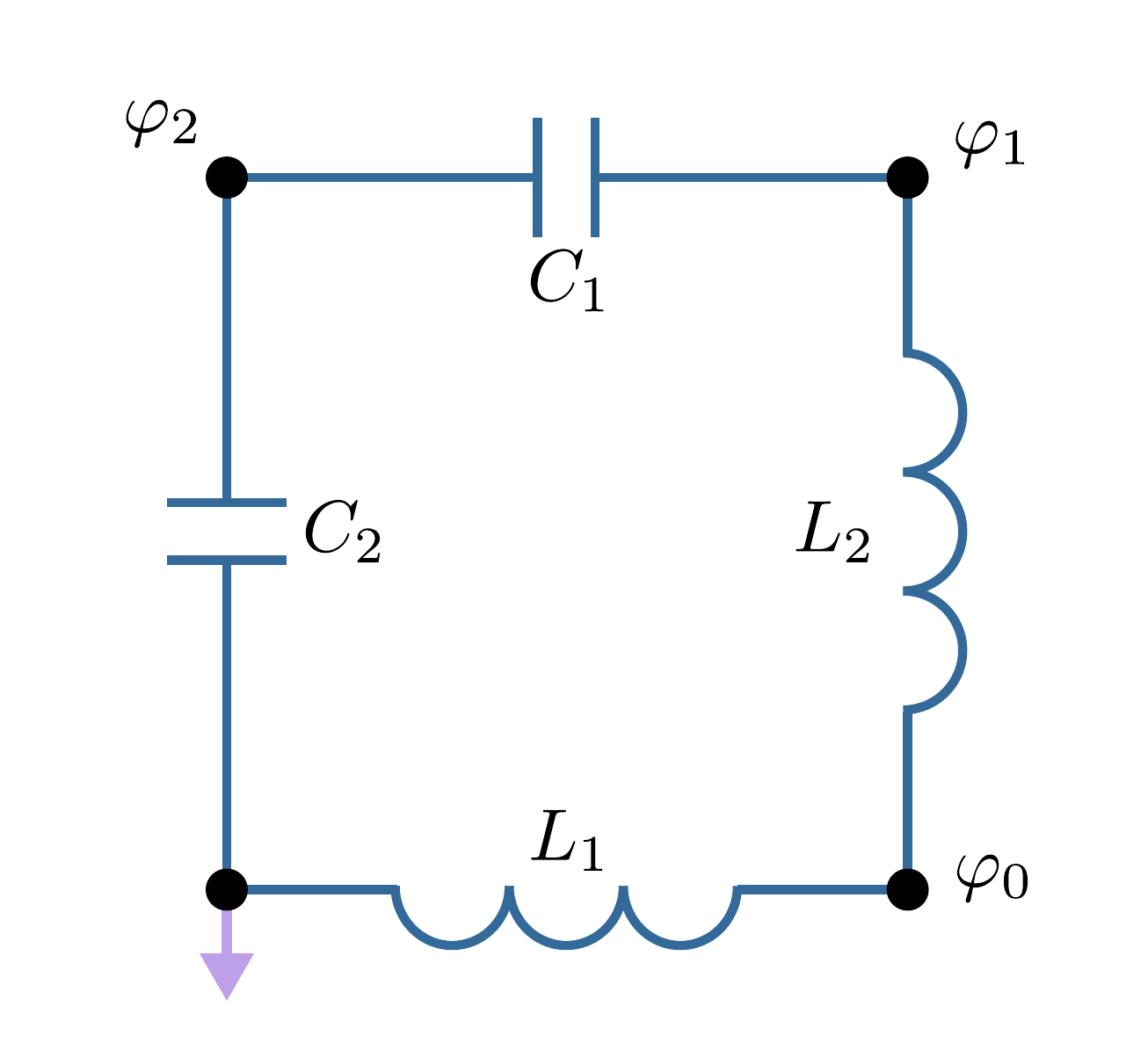}
\caption{(Color online) Schematic illustration of a building block containing passive nodes. The system consists of two series-connected inductors $L_{1}$ and $L_{2}$, and two capacitors $C_{1}$ and $C_{2}$.  }
\label{fig:figA1}
\end{figure}
In this section, we illustrate how to eliminate the passive nodes present in a random configuration of the single loop. For doing so, let us consider the circuit depicted in Fig. (\ref{fig:figA1}) whose Lagrangian reads
\begin{eqnarray}
	\label{EqnA1}\nonumber
	\mathcal{L} &=& \bigg(\frac{\Phi_{0}}{2\pi}\bigg)^{2}\bigg[\frac{C_{1}(\dot{\varphi}_{2}-\dot{\varphi}_{1})^{2}}{2} + \frac{C_{2}\dot{\varphi}_{2}^{2}}{2}\\
	 &-& \frac{\varphi_{0}^{2}}{2L_{1}} - \frac{(\varphi_{1}-\varphi_{0})^2}{2L_{2}}\bigg].
\end{eqnarray}
As no Josephson junction exists in this configuration, we do not apply the fluxoid quantization rule. Afterward, we calculate the Euler-Lagrange equations for all the node variables present in the configuration, obtaining
\begin{subequations}
	\begin{eqnarray}
		\label{EqnA2a}\nonumber
		\frac{\partial \mathcal{L}}{\partial[\dot{\varphi}_{0}]} &=& 0,\quad\frac{\partial \mathcal{L}}{\partial[\varphi_{0}]}=-\frac{\varphi_{0}}{L_{1}}+\frac{(\varphi_{1}-\varphi_{0})}{L_{2}},\\\\
		\label{EqnA2b}\nonumber
		\frac{\partial \mathcal{L}}{\partial[\dot{\varphi}_{1}]} &=& -C_{1}(\dot{\varphi}_{1}-\dot{\varphi}_{0}),\quad\frac{\partial \mathcal{L}}{\partial[\varphi_{1}]}=-\frac{(\varphi_{1}-\varphi_{0})}{L_{2}},\\\\
		\label{EqnA2c}\nonumber
		\frac{\partial \mathcal{L}}{\partial[\dot{\varphi}_{2}]} &=& C_{1}(\dot{\varphi}_{2}-\dot{\varphi}_{1})+C_{2}\dot{\varphi}_{2},\quad\frac{\partial \mathcal{L}}{\partial[\varphi_{2}]}=0.\\
	\end{eqnarray}
\end{subequations}
From these equations, it is easy to note that some generalized coordinates and velocities are missing. In particular, Eq.~(\ref{EqnA2a}) lacks generalized velocity, and Eq.~(\ref{EqnA2c}) misses the generalized coordinate. These variables constitute the passive nodes of the circuit, and we can eliminate them by solving the non-zero part of the Euler-Lagrange equation in terms of the active node, namely $\varphi_{1}$, obtaining
\begin{subequations}
	\begin{eqnarray}
		\label{EqnA3a}
		\varphi_{0} &=& \bigg(\frac{L_{1}}{L_{1}+L_{2}}\bigg)\varphi_{1},\\
		\label{EqnA3b}
		\dot{\varphi}_{2}&=&\bigg(\frac{C_{1}}{C_{1}+C_{2}}\bigg)\dot{\varphi}_{1}.
	\end{eqnarray}
\end{subequations}
\begin{figure*}[!t]
	\centering
	\includegraphics[width=1\linewidth]{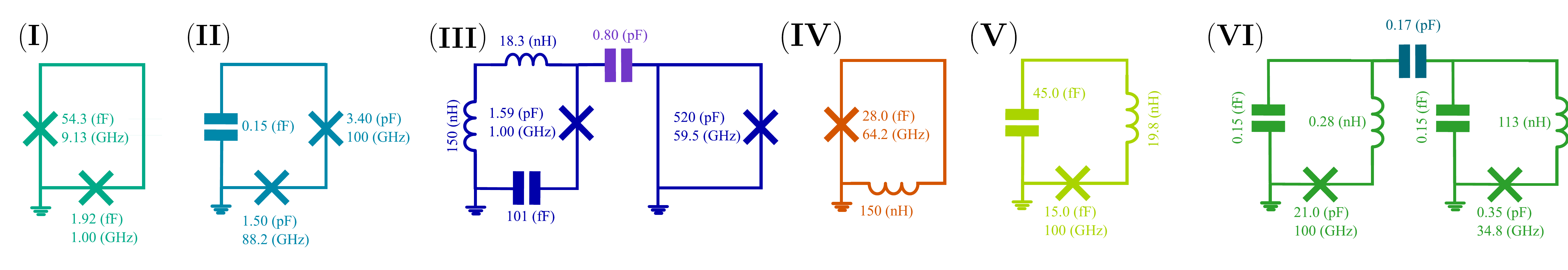}
	\caption{(Color online) Illustration of the optimal circuit configuration obtained in Table~\ref{Table1} and Table~\ref{Table2}. For the Josephson junctions, we show the values of the capacitances in Farad, whereas the Josephson energy is presented in frequency units.}
	\label{fig:figA4}
\end{figure*}
We obtain the circuit Lagrangian without passive nodes just by replacing Eq.~(\ref{EqnA3a}) and Eq.~(\ref{EqnA3b}) in the Lagrangian in Eq.~(\ref{EqnA1})
\begin{eqnarray}
	\label{EqnA4}
	\mathcal{L} = \bigg(\frac{\Phi_{0}}{2\pi}\bigg)^{2}\bigg[\bigg(\frac{C_{1}C_{2}}{C_{1}+C_{2}}\bigg)\frac{\dot{\varphi}_{1}^{2}}{2}-\frac{\varphi_{1}^2}{2(L_{1}+L_{2})}\bigg].
\end{eqnarray}
Notice that now, the circuit Lagrangian is described by a single degree of freedom that corresponds to an LC circuit with effective capacitance and inductance, respectively.\par
\begin{figure}[!t]
	\centering
	\includegraphics[width=1\linewidth]{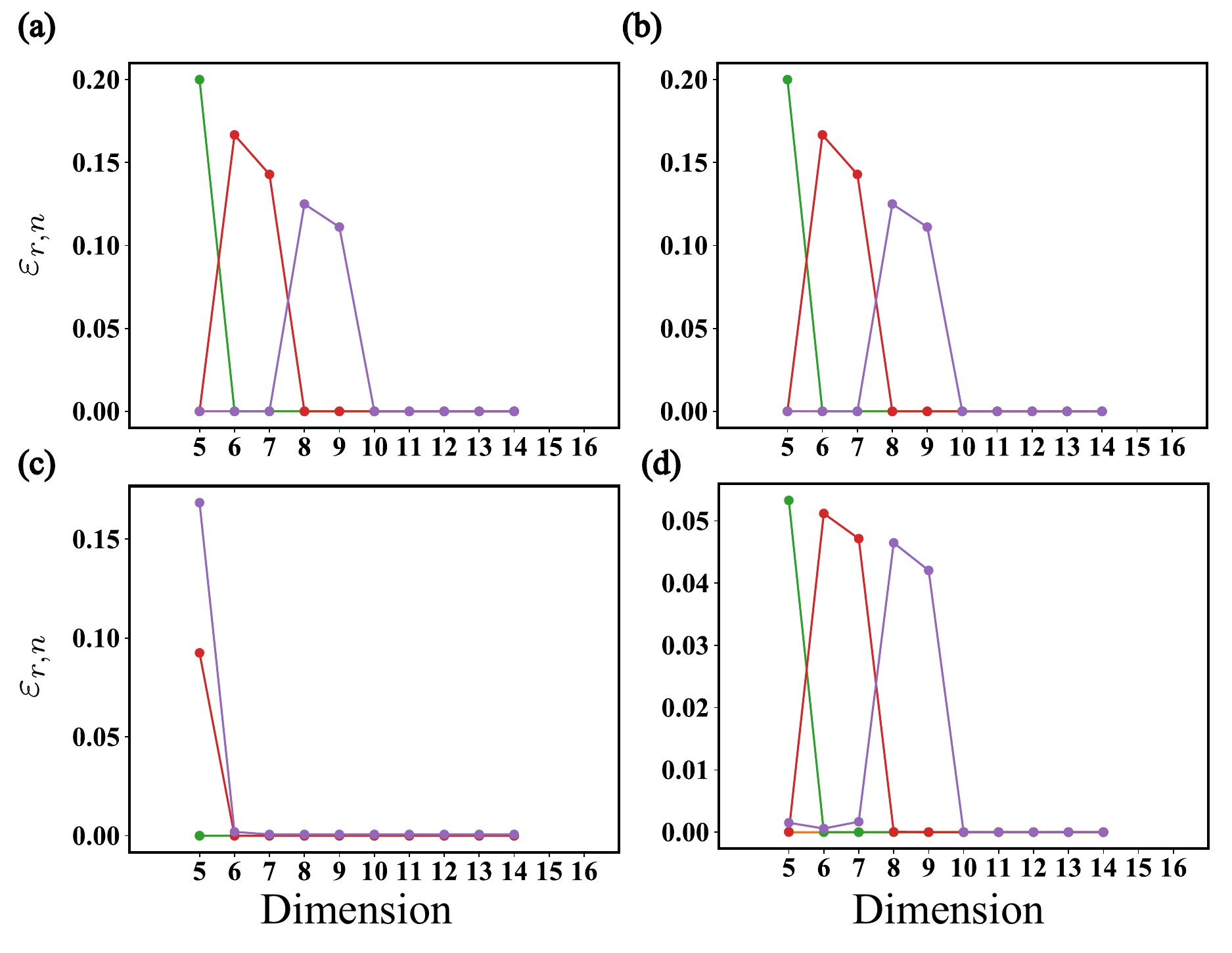}
	\caption{(Color online) Relative error $\varepsilon_{r,n}$ for the four low-lying energy levels for the following circuit configuration; (a) $\vec{\alpha}=\{1,0,0,0\}$, $\vec{\beta}=\{0,0,0,1\}$, $\vec{\gamma}=\{0,0,0,0\}$, (b) $\vec{\alpha}=\{1,0,0,0\}$, $\vec{\beta}=\{0,0,0,0\}$, $\vec{\gamma}=\{0,0,0,1\}$, (c) $\vec{\alpha}=\vec{\beta}=0$, $\vec{\gamma}=\{1,1,1,0\}$, and (d) $\vec{\alpha}=\{1,0,0,0\}$, $\vec{\beta}=\{0,1,0,0\}$, $\vec{\gamma}=\{0,0,1,0\}$. The system parameter are given by $C=C_{J}=10~ \rm{fF}$, $L=645~\rm{nH}$, and $E_{J}=10^{10}\hbar$. }
	\label{fig:figA2}
\end{figure}
Now, both kinetic and inductive terms contain the same variables. Notice that this Lagrangian is the same as considering a circuit with only one inductance with a value equal to the effective inductance of the passive node. 
\section{Numerical Convergence of the Single loop circuit}
In this section, we perform a numerical analysis concerning the energy spectrum of the single-loop Hamiltonian given in Eq. (\ref{Eqn11}). We focus on two aspects; the convergence of Hilbert space describing the single loop circuit Hamiltonian and how the non-linear terms change the anharmonicity in the energy spectrum.
\subsection{Optimal Hilbert space size}\label{converge_hilber}
We compute the energy spectrum of the harmonic part of Hamiltonian from Eq. (\ref{Eqn11}), and we diagonalize it by increasing the Hilbert space by adding one excitation on each mode. Then, we compare the $n$th energy level for the system containing $m+1$ and $m$ excitation per mode through the relative error defined as follows 
\begin{eqnarray}
 \varepsilon_{r,n}=\left|\frac{\epsilon_{n}(m)-\epsilon_{n}(m+1)}{\epsilon_{n}(m+1)}\right|.
 \end{eqnarray}
Here, $\epsilon_{n}(m)$ corresponds to the $n$th energy level computed with $m$ excitations in each mode. Fig (\ref{fig:fig2}) shows the relative error $\varepsilon_{r,n}$ of the first four energy levels to four different circuit configurations, where each of them is characterized by a specific triplet $\{\vec{\alpha},\vec{\beta},\vec{\gamma}\}$. The numerical diagonalization of the Hamiltonian for these different configurations shows that for an $m=12$, the relative error between the $n$th energy levels is closest to zero. Thus, we choose that $m$ as our optimal Hilbert space size.
\section{Illustration optimal circuit}
In this section, we show the optimal circuit configurations obtained with our algorithm for the ladder and lambda configuration depicted in Table~\ref{Table1} and Table~\ref{Table2}

\section{Comparison between different optimization algorithms}
\begin{figure}
    \centering
    \includegraphics[width=1\linewidth]{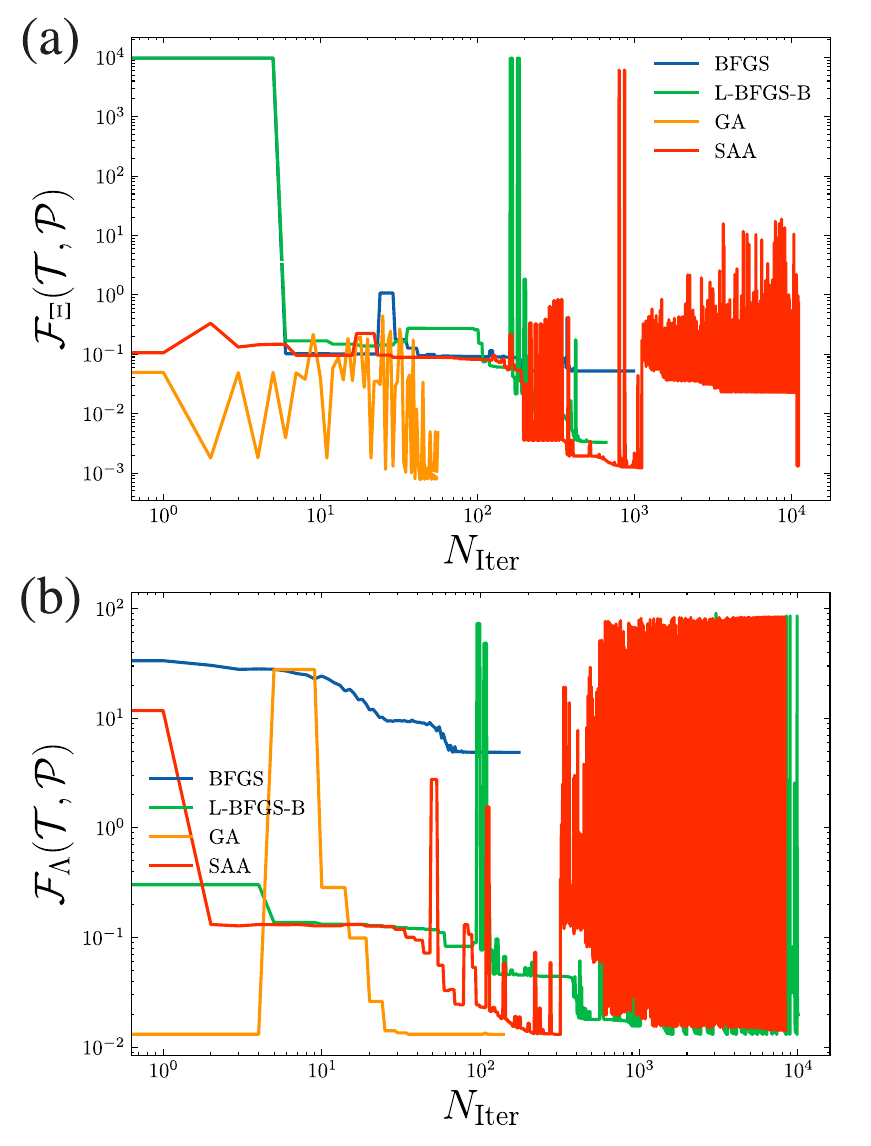}
    \caption{Comparison of different optimizers for finding the optimal circuit parameters of the selected topology for (a) the $\Xi$ system and (b) the $\Lambda$ system. Gradient-based optimization algorithms fail to converge reliably in both cases, showing abrupt variations in the cost function and becoming trapped in local minima. In contrast, simulated annealing achieves better convergence in the early iterations but exhibits persistent oscillations and high variance across steps. The genetic algorithm outperforms the others, reaching lower cost function values in fewer iterations. Although some fluctuations are still present, they are primarily due to the mutation operations during evolution. Moreover, the solutions found by the genetic algorithm demonstrate robustness to perturbations in circuit parameters, making the approach well-suited for experimental implementations that must account for fabrication errors.}
    \label{fig:FigA4}
\end{figure}

\end{document}